\pdfoutput=1
\newcommand*{\ATLASLATEXPATH}{}
\documentclass[UKenglish,texlive=2016,PAPER,cernpreprint]{\ATLASLATEXPATH atlasdoc}
\makeatletter
\DeclareOldFontCommand{\rm}{\normalfont\rmfamily}{\mathrm}
\DeclareOldFontCommand{\sf}{\normalfont\sffamily}{\mathsf}
\DeclareOldFontCommand{\tt}{\normalfont\ttfamily}{\mathtt}
\DeclareOldFontCommand{\bf}{\normalfont\bfseries}{\mathbf}
\DeclareOldFontCommand{\it}{\normalfont\itshape}{\mathit}
\DeclareOldFontCommand{\sl}{\normalfont\slshape}{\@nomath\sl}
\DeclareOldFontCommand{\sc}{\normalfont\scshape}{\@nomath\sc}
\makeatother

\usepackage[subcaption,backend=biber]{\ATLASLATEXPATH atlaspackage}

\usepackage{\ATLASLATEXPATH atlasbiblatex}
\DeclareFieldFormat[article]
{pages}{#1}
\DeclareFieldFormat[Article]
{pages}{#1}

\usepackage{\ATLASLATEXPATH atlascontribute}

\usepackage[BSM,process]{\ATLASLATEXPATH atlasphysics}

\usepackage{lscape}
\graphicspath{{logos/}{figures/}}

\usepackage{PaperDM-defs}

\usepackage{footmisc}

\AtlasTitle{Search for dark matter produced in association with bottom or top quarks in $\sqrt{s}=13$~TeV $pp$ collisions with the ATLAS detector}

\author{The ATLAS Collaboration}

\AtlasRefCode{SUSY-2016-18}

\PreprintIdNumber{CERN-EP-2017-229}

\AtlasJournalRef{Eur. Phys. J. C 78 (2018) 18}
\AtlasDOI{10.1140/epjc/s10052-017-5486-1}

\AtlasAbstract{A search for weakly interacting massive dark-matter particles produced in association with bottom or top quarks is presented.
Final states containing third-generation quarks and missing transverse momentum are considered.
The analysis uses \intlumi\ of proton--proton collision data recorded by the ATLAS experiment at $\sqrt{s}=13$~TeV in 2015 and 2016.
No significant excess of events above the estimated backgrounds  is observed.
The results are interpreted in the framework of simplified models
of spin-0 dark-matter mediators. 
For colour-neutral spin-0 mediators produced in association with top
quarks and decaying into a pair of dark-matter particles, mediator
masses below 50~GeV are excluded assuming a dark-matter candidate mass
of 1~GeV and unitary couplings. 
For scalar and pseudoscalar mediators produced in association with
bottom quarks, the search sets limits on  
the production cross-section 
 of $300$ times the predicted rate for mediators with masses between $10$
 and $50\; \GeV$ and assuming a dark-matter mass of $1 \;\GeV$ and unitary coupling.
Constraints on colour-charged scalar simplified models are also
presented. Assuming a dark-matter particle mass of $35\; \GeV$, mediator particles
with mass below $1.1\; \TeV$ are excluded for couplings yielding a dark-matter relic density consistent with measurements.
}

\hypersetup{pdftitle={ATLAS document},pdfauthor={The ATLAS Collaboration}}

\begin{document}
\maketitle

\section{Introduction}
\label{sec:intro}

Astrophysical observations have provided compelling evidence for the
existence of a non-baryonic dark component of the universe:
dark matter (DM)~\cite{Zwicky:1933gu, DMreview}.
The currently most accurate, although somewhat indirect, determination of DM abundance comes from global
fits of cosmological parameters to a variety of observations~\cite{Komatsu:2010fb,Ade:2015xua},
while the nature of DM remains largely unknown.
One of the
candidates for a DM particle is a weakly interacting massive particle
(WIMP)~\cite{Steigman:1984ac}.
At the Large Hadron Collider (LHC), one can search for WIMP DM ($\chi$) pair  production
in \emph{pp} collisions.
WIMP DM would not be detected
and
its production leads to signatures with missing transverse momentum.
Searches for the production of DM in association with Standard Model (SM) particles
have been performed
at the LHC \cite{CMS-EXO-16-012,CMS-EXO-16-037,Sirunyan:2017ewk,Aaboud:2017uak,EXOT-2016-32,EXOT-2015-03,EXOT-2016-25}.

Recently proposed simplified benchmark models for DM
production assume the existence of a mediator particle which couples
both to the SM and to the dark sector~\cite{Abercrombie:2015wmb,Buckley:2014fba,Haisch:2015ioa}.
The searches presented in this paper focus on the case of a fermionic DM particle
produced through the exchange of a spin-0 mediator, which can be
either a colour-neutral scalar or pseudoscalar particle (denoted by $\phi$ or $a$, respectively) or a
colour-charged scalar mediator ($\phi_b$).
The couplings of the mediator to the SM fermions are
severely restricted by precision flavour
measurements. An ansatz that automatically
relaxes these constraints is Minimal Flavour Violation~\cite{DAmbrosio:2002vsn}. This
assumption implies that the interaction between any new neutral spin-0
state and SM matter is proportional to the fermion masses via
Yukawa-type couplings\footnote{Following Ref.~\cite{Buckley:2014fba}, couplings to $W$ and $Z$ bosons, as well as explicit dimension-4 $\phi$--$h$ or $a$--$h$ couplings, are set to zero in this simplified model.
In addition, the coupling of the mediator to the dark sector are not taken to be proportional to the mass of the DM candidates.}.
It follows that colour-neutral mediators would be sizeably
produced through loop-induced gluon fusion or in association with heavy-flavour quarks.
The characteristic signature used to
search for the former process
is a high transverse momentum jet recoiling
against missing transverse momentum~\cite{EXOT-2015-03,CMS-EXO-16-037}.

\begin{figure}[b]
\centering
\begin{subfigure}{.28\textwidth}
\includegraphics[width=.98\textwidth]{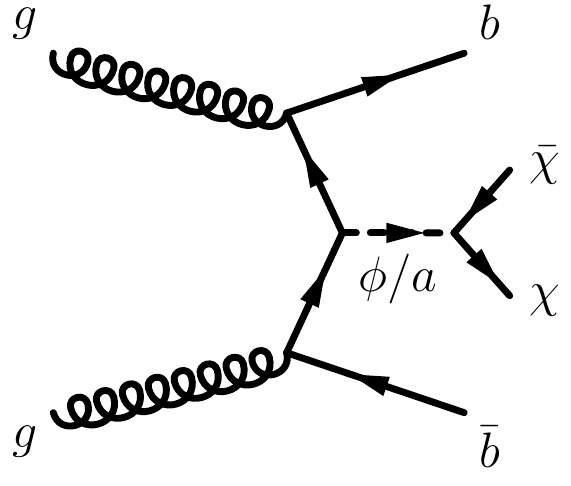}
\caption{}\label{fig:Spin0_feynman}
\end{subfigure}
\hspace{25pt}
\begin{subfigure}{.28\textwidth}
\includegraphics[width=.98\textwidth]{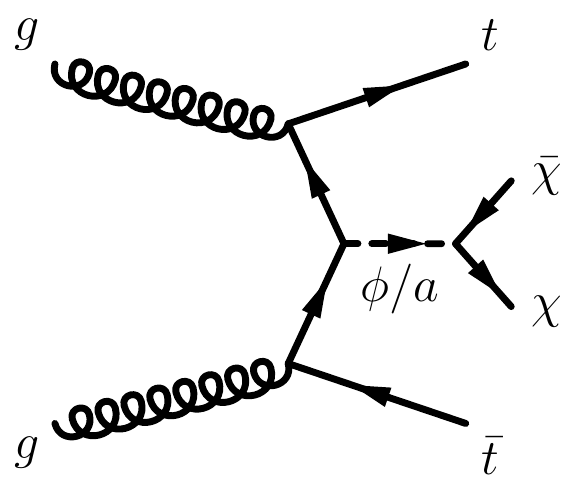}
\caption{}\label{fig:Spin0ttbar}
\end{subfigure}
\hspace{25pt}
\begin{subfigure}{.28\textwidth}

\vspace{-6pt}
\includegraphics[width=.98\textwidth]{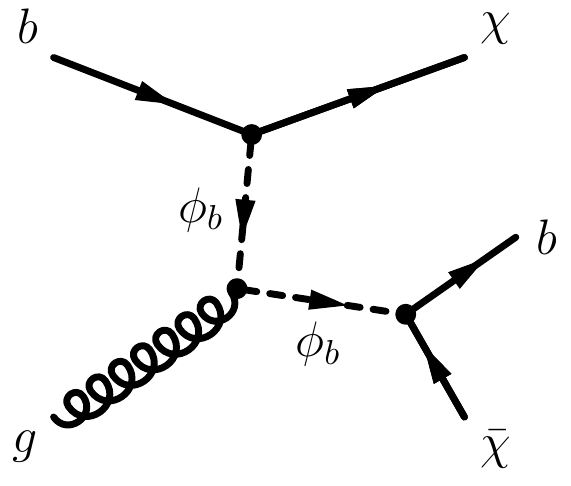}
\caption{}
\label{fig:bFDM_feynman}
\end{subfigure}
\caption{Representative diagrams at the lowest order for spin-0 mediator associated production with top and bottom quarks:~\subref{fig:Spin0_feynman}
colour-neutral spin-0 mediator associated production with bottom quarks \bbbar+$\phi/a$;
\subref{fig:Spin0ttbar} colour-neutral spin-0 mediator associated production with top quarks \ttbar+$\phi/a$;
\subref{fig:bFDM_feynman} colour-charged scalar mediator model decaying into a bottom quark and a DM particle \bFDM .}
\label{fig:feynmans}
\end{figure}

This paper focuses on dark matter produced in association with heavy flavour (top and bottom) quarks.
These final states were addressed by the CMS Collaboration in Ref.~\cite{Sirunyan:2017xgm}.
For signatures with two top quarks (\ttbar+$\phi/a$),
final states where both $W$ bosons decay into hadrons or both $W$ bosons decay into leptons
are considered in this paper. They are referred to as fully hadronic and dileptonic \ttbar\ decays, respectively.
Searches in final-state events characterised by fully hadronic or dileptonic
top-quark pairs have been carried out targeting supersymmetric partners of
the top quarks~\cite{Stop0L,Stop2L}. Due to the different kinematics of
the events under study, those searches are not optimal for the DM
models considered in this paper.
The search in the channel where one $W$ boson decays into hadrons and one $W$ boson decays into leptons (semileptonic \ttbar\ decays)
is presented together with the searches for top squarks in the same channel~\cite{Stop1L}.
Signatures with bottom quarks in the final state are denoted \bbbar+$\phi/a$ in the following.
Representative diagrams for tree-level production of these
models are shown in Figs.~\ref{fig:Spin0_feynman}~and~\ref{fig:Spin0ttbar}.
Processes with similar kinematic properties might also occur in two-Higgs-doublet models~\cite{2HDM}.
Following the notation of Ref.~\cite{Buckley:2014fba}, the model
has four parameters: the mass of the mediator $m_\phi$ or $m_a$,
the DM mass $m_\chi$,  the DM--mediator coupling $g_\chi$, and the flavour-universal SM--mediator
coupling $g_q$.
The mediator width is assumed to be the minimal width, which is the one calculated
from the masses and couplings assumed by the model~\cite{Abercrombie:2015wmb}.
The mediator can decay into SM particles or into DM particles.
This search is sensitive to decays
of the mediator into a pair of DM particles.
Off-shell DM production is also taken into account.
The effective production cross-section of DM particles at $pp$ colliders is a function of the
production cross-section of the mediator, depending on $g_q$, and on the
branching ratio for the mediator to decay into a pair of DM particles, which is a function of $g_q$ and $g_\chi$ \cite{Abercrombie:2015wmb}.
The cross-section for DM production is therefore proportional to the squared product of the couplings ($g_q\;\cdot g_\chi$)$^2$,
and an additional assumption of $g_q = g_\chi = g$ is made to reduce the number of parameters.
Since the cross-section of annihilation and scattering from nucleons has the same functional dependence on the couplings, the
same assumption is made when the results are compared to non-collider experiments.

The second category of models considered in this search
is the case of colour-charged  scalar mediators~\cite{Agrawal:2014una}.
The model assumes bottom-flavoured dark matter (\bFDM) and
was proposed to explain the excess of gamma rays from the galactic centre observed by the
Fermi Gamma-ray Space Telescope, if this excess is to be
interpreted as
a signal for DM annihilation \cite{Daylan:2014rsa}, while alternative conjectures without DM are also discussed~\cite{Fermi:1705}.
A representative diagram for the production of this signal is shown in Fig.~\ref{fig:bFDM_feynman}.
In this model, a new scalar field, $\phi_b$, mediates the interaction between
DM and quarks. Dark matter is assumed to be the lightest Dirac fermion that belongs to a flavour-triplet coupling to right-handed, down-type quarks.
The cosmological DM is the third component of the triplet and couples preferentially to bottom quarks.
It explains the galactic-centre excess if a DM mass around $35\;\GeV$ is assumed.
The other Dirac fermions in the flavour-triplet are heavy and couple weakly, and are therefore neglected.
The \bFDM\ model has three parameters: the mediator and the DM masses ($m(\phi_b)$ and $m(\chi)$, respectively),
and the coupling strength between the mediator and the DM particle,  $\lambda_b$~\cite{Agrawal:2014una}.
For each pair of mass values considered, $\lambda_b$ is set to the value, generally larger than one, predicting a DM relic density compatible with
the astrophysical observations as detailed in Ref.~\cite{Agrawal:2014una}.
Strong-interaction pair production of $\phi_b$, which does not depend on the coupling,
is equivalent to the pair production of the lightest supersymmetric partner of the bottom quark
(bottom squark, $\tilde{b}_1$) assuming that it decays exclusively into a bottom quark and the lightest
neutralino ($\tilde{\chi}^0_1$). Exclusion limits on $m(\tilde{b}_1)$,  which depend on $m(\tilde{\chi}^0_1)$, are set in
dedicated searches by the ATLAS and CMS collaborations~\cite{Sbottom, Sirunyan:2016jpr}.
The target of this search is the single production mode represented in Fig.~\ref{fig:bFDM_feynman},
which can dominate the
production rate of the $\phi_b$ mediator due to the relatively large
values assumed for $\lambda_b$.
The parameter space considered corresponds to $\phi_b$ masses of a few hundred GeV.
A search by the ATLAS Collaboration with the $\sqrt{s} = 8\; \TeV$ LHC Run-1 dataset
has already excluded $m(\phi_b) < 600\;\GeV$ for $m(\chi) = 35\;\GeV$ \cite{Aad:2014vea}.

Four experimental signatures are considered in this paper.
The first two signatures consist of event topologies with large missing transverse momentum and either one or
two bottom quarks, while the other two consist of events with large missing transverse momentum and two top
quarks, decaying either dileptonically or fully hadronically.
The search presented in this paper is based on a set of independent analyses optimised for these four
experimental signatures and searches for dark-matter production via colour-charged and colour-neutral mediators.

\section{Detector description and event reconstruction}
\label{sec:detector}
The ATLAS experiment~\cite{ATLASdetector} is a multi-purpose particle detector with a forward-backward symmetric cylindrical
geometry and nearly $4\pi$ coverage in solid angle.\footnote{ATLAS uses
a right-handed coordinate system with its origin at the nominal
interaction point (IP) in the centre of the detector and the
$z$-axis along the beam pipe. The $x$-axis points from the IP to the
centre of the LHC ring, and the $y$-axis points upward. Cylindrical
coordinates ($r$, $\phi$) are used in the transverse plane, $\phi$
being the azimuthal angle around the beam pipe. The pseudorapidity
is defined in terms of the polar angle $\theta$ as $\eta = -\ln
\tan(\theta/2)$. Rapidity is defined as $y=0.5 \ln\left[(E + p_z
)/(E - p_z )\right]$ where $E$ denotes the energy and $p_z$ is the
component of the momentum along the beam direction.}
It consists of an inner tracking detector (ID) surrounded by a
superconducting solenoid, electromagnetic and hadronic calorimeters,
and an external muon spectrometer incorporating large superconducting
toroidal magnets.
The inner tracking detector consists of pixel and silicon microstrip detectors
covering the pseudorapidity region $|\eta|<2.5$, surrounded by a transition radiation tracker
which provides electron identification in the region $|\eta|<2.0$.
Between Run 1 and Run 2, a new inner pixel layer, the insertable
B-layer~\cite{IBL}, was inserted at a mean sensor radius of
$3.3$~cm. The inner detector is surrounded by a thin superconducting
solenoid providing an axial $2$~T magnetic field and by a
fine-granularity lead/liquid-argon (LAr) electromagnetic calorimeter
covering $|\eta|<3.2$. A steel/scintillator-tile calorimeter provides
hadronic coverage in the central pseudorapidity range ($|\eta|<1.7$).
The end-cap and forward regions ($1.5<|\eta|<4.9$) of the hadronic
calorimeter are made of LAr active layers with either copper or
tungsten
as the absorber material.
A muon spectrometer with an air-core toroid magnet system surrounds the calorimeters.
Three stations of high-precision tracking chambers provide coverage in
the range $|\eta|<2.7$, while dedicated chambers allow triggering
in the region $|\eta|<2.4$. The ATLAS trigger system consists of a
hardware-based level-1 trigger followed by a software-based high-level
trigger~\cite{TRIG-2016-01}.

The events used in this analysis are required to pass
either an online trigger requiring a minimum of two electrons, two muons or an electron and a
muon, or an online missing transverse momentum trigger selection.
The trigger thresholds are such that a plateau of the efficiency
is reached for events passing the
analysis requirements presented in Sect.~\ref{sec:selection}.
The events are also required to have a reconstructed
vertex~\cite{ATL-PHYS-PUB-2015-026} with at least two associated
tracks with transverse momentum (\pT) larger than $400\;\MeV$ which are
consistent with originating from the beam collision region.
The vertex with the highest scalar sum of the squared
transverse momenta of the associated tracks is considered to be the
primary vertex of the event.

This analysis requires the reconstruction of jets, muons,
electrons, photons and missing transverse momentum.
Jets are reconstructed from three-dimensional energy clusters in the
calorimeter~\cite{PERF-2014-07} using the anti-$k_t$ jet
clustering algorithm~\cite{Cacciari:2008} with a radius parameter
$R=0.4$ implemented in the FastJet package~\cite{Cacciari:2011ma}. Jets are calibrated
as described in
Ref.~\cite{PERF-2016-04}, and the
expected average energy contribution from
clusters resulting from additional $pp$ interactions in the same or nearby bunch crossings (pile-up interactions) is
subtracted according to the jet area~\cite{ATL-PHYS-PUB-2015-015}.
Only jet candidates (baseline jets) with $\pT>20\;\GeV$ and $|\eta|<2.8$
are considered in the analysis.
Quality criteria identify jets arising from non-collision sources
or detector noise and any event containing such a jet is removed~\cite{ATLAS-CONF-2015-029,ATLAS-CONF-2010-038}.
Additional selection requirements are
imposed on jets with $\pT < 60\;\GeV$ and $|\eta|<2.4$ in order to reject jets produced in pile-up interactions~\cite{PERF-2014-03}.
Jets are also reclustered into larger-radius jets ($R=0.8$ or $1.2$) by
applying the anti-$k_t$ clustering algorithm to the
$R=0.4$ jets. These jets are exploited
to identify $W$-boson decays into a pair of quarks and also to identify top-quark
candidates.

Jets containing $b$-hadrons (\bjets) and which are within the inner detector
acceptance ($|\eta|<2.5$) are identified (\btagged) with a multivariate algorithm
that exploits
the impact parameters of the charged-particle tracks, the presence of secondary
vertices and the reconstructed flight paths of $b$- and $c$-hadrons inside
the jet~\cite{PERF-2012-04,ATL-PHYS-PUB-2016-012}.
Depending on the signal region requirements detailed in Sect.~\ref{sec:selection}, a "medium" or  "tight"  working-point is
used for the \bjet\  identification, corresponding to an average
efficiency for $b$-quark jets in simulated \ttbar\ events of 77\% and
60\%, respectively. An additional "loose" working-point with 85\%
efficiency for $b$-quark jets in simulated \ttbar\ events is used to resolve
ambiguities in the reconstruction of physics objects, as described at the end of this section.

Muon candidates are reconstructed in the region $|\eta|<2.7$ from muon
spectrometer tracks matching ID tracks (where applicable).
The pseudorapidity requirements are restricted to $|\eta|<2.4$ for events
passing the muon online trigger criteria, due to the coverage of the
muon triggering system.
Events containing one or more
muon candidates that have a transverse (longitudinal) impact parameter with respect to the primary vertex
larger than $0.2$ mm ($1$ mm) are rejected to suppress muons from cosmic rays.
Baseline candidate muons, used for the definition of vetoes in all signal regions but those searching for fully hadronic top decays,
must have $\pT>\SI{10}{\GeV}$ and pass the "medium"
identification requirements defined in Ref.~\cite{PERF-2015-10}.
The baseline candidate muons used in fully hadronic $\ttbar$ final states
are instead required to pass the "loose" identification requirements~\cite{PERF-2015-10} and to have $\pT>\SI{6}{\GeV}$,
in order to strengthen the veto definition.
Baseline electron candidates are reconstructed from isolated electromagnetic
calorimeter energy deposits matched to ID tracks and are required to
have $|\eta|<2.47$ and $\pT>\SI{10}{\GeV}$, and must
pass a "loose" likelihood-based identification
requirement~\cite{PERF-2016-01,ATL-PHYS-PUB-2015-041}.

Stricter requirements are imposed on the baseline lepton (electron or muon) definitions for the selection criteria requiring leptons in the final state.
Signal muon candidates, used for all selection requirements with leptons in the final state,
must have $\pt > 20\;\GeV$ and satisfy "medium"
identification criteria~\cite{PERF-2015-10}. Furthermore, they are required
to be isolated using a "loose" criterion designed to be
99\% efficient for muons from $Z$-boson decays~\cite{PERF-2015-10}.
Signal electron candidates are required to pass "tight" requirements on the
likelihood-based identification~\cite{PERF-2016-01} and must have $\pt > 20\;\GeV$.
In order to improve signal acceptance,
the requirement on the likelihood-based identification is relaxed to "medium" for the signal region optimised for the two-lepton final state.
Like the muons, signal electrons are required
to be isolated from other activity using a "loose" isolation criterion~\cite{ATLAS-CONF-2016-024}.
Signal electrons (muons) are matched to the primary vertex (PV) of the event (see Sect.~\ref{sec:selection}) by
requiring their transverse impact parameter $d^{\mathrm P \mathrm V}_0$, with respect to the primary vertex, to
have a significance $|d^{\mathrm P \mathrm V}_0/\sigma(d^{\mathrm P \mathrm V}_0)| < 5\; (3)$.
In addition, for both the electrons and muons
the longitudinal impact parameter $z^{\mathrm P \mathrm V}_0$ and the polar angle $\theta$
are required to satisfy $|z^{\mathrm P \mathrm V}_0 \sin\theta| < 0.5$ mm.
In the following, the combination of signal electrons and muons optimised for the
two-lepton final state is referred to as the medium-lepton requirement.
Similarily, the combination of the signal electrons and muons passing the "tight" identification criteria is referred to as the
tight-lepton requirement.
The number of leptons passing the medium and tight requirements
is denoted by $\mathcal{N}^{\mathrm M}_{\ell}$ and $\mathcal{N}^{\mathrm T}_{\ell}$, respectively.

Photons are reconstructed from clusters of energy deposits in the electromagnetic calorimeter measured
in projective towers~\cite{PERF-2013-05,PERF-2013-04}.
Photon candidates are required to have $\pT > 10\; \GeV$ and $|\eta|<2.37$, whilst being outside the transition
region $1.37 < |\eta| < 1.52$ between the barrel and end-cap calorimeters, and to satisfy "tight" identification criteria~\cite{PERF-2013-04}. The
photons used in this analysis are further required to have $\pT > 130\; \GeV$ and to be isolated~\cite{PERF-2013-05}.

To resolve reconstruction ambiguities, an overlap removal algorithm is applied to
loose candidate leptons and jets.
Jet candidates with $\pT>20\;\GeV$ and $|\eta|<2.8$
are removed if they are not $b$-tagged when employing the loose working-point and are within $\Delta R
=\sqrt{(\Delta y)^2+(\Delta\phi)^2} = 0.2$ of an electron candidate.
The same is done
for jets which lie close to a muon candidate and have less than three associated tracks or a ratio of
muon \pt\ to jet \pt greater than 0.5. Finally,
any lepton candidate within $\Delta R = 0.4$ of the direction of a
surviving jet candidate is removed, in order to reject leptons from
the decay of a $b$- or $c$-hadron. Electrons which share an ID track
with a muon candidate are also removed.

The missing transverse momentum vector, \ptmiss, whose magnitude is denoted
by \MET, is defined as the negative vector sum of the transverse
momenta of all identified physics objects (electrons, photons, muons,
jets) and an additional soft term. The soft term is constructed from
all tracks that originate from the primary vertex but are not
associated with any physics object. In this way, the \MET is adjusted
for the calibration of the jets and the other identified physics
objects above, while maintaining pile-up independence in the soft
term~\cite{ATL-PHYS-PUB-2015-023,ATL-PHYS-PUB-2015-027}.

\section{Data and simulated event samples}
\label{sec:dataMC}
The dataset used in this analysis consists of $pp$ collision data
recorded at a centre-of-mass energy of $\sqrt{s} = 13~\TeV$ with stable beam
conditions.
The integrated
luminosity of the combined 2015+2016
dataset after requiring that all detector subsystems were operational during data recording is \intlumi.
The uncertainty in the total integrated luminosity is 3.2\%,
derived following a methodology similar to that detailed in Ref.~\cite{DAPR-2013-01}.

Monte Carlo (MC) simulated event samples are used to aid in the estimation of the background from SM processes and to model the dark-matter signal.
All simulated events were processed through an ATLAS detector simulation~\cite{Aad:2010ah} based on {\sc Geant4}~\cite{Agostinelli:2002hh} or
through a fast simulation using a parameterisation of the calorimeter response and {\sc Geant4} for the other parts of the detector~\cite{ATL-PHYS-PUB-2010-013}.
The simulated events are reconstructed with the same reconstruction algorithms used for data.
Correction factors are applied to the simulated events to compensate for differences between data and MC simulation in the
$b$-tagging efficiencies and mis-tag rates,
lepton and photon identification, reconstruction and trigger efficiencies.
The  MC samples are reweighted so that the pile-up distribution matches the one observed in the data.

The matrix element (ME) generator, parton shower (PS), cross-section normalisation, parton distribution function (PDF) set and the set of tuned parameters (known as tune) describing the underlying event for these samples are given in Table~\ref{tab:MC}, and more details of the generator configurations can be found in Refs.~\cite{ATL-PHYS-PUB-2016-004,ATL-PHYS-PUB-2016-003,ATL-PHYS-PUB-2016-002,ATL-PHYS-PUB-2016-005}.
The generation of \ttbar pairs and single-top-quark processes in the $Wt$- and $s$-channels was performed using the \POWHEGBOX
v2 generator with the CT10 PDF set for the matrix element calculations. Electroweak
$t$-channel single-top-quark events were generated using the \POWHEGBOX v1 generator.
For all processes, a top-quark mass of 172.5 GeV is assumed.
The PS and the underlying event were simulated using \PYTHIA
6.428 with the CT10 PDF set.
Samples of single-top-quark and \ttbar production are normalised
to their NNLO cross-section including the resummation of soft gluon emission
at next-to-next-to-leading-log (NNLL) accuracy using \textsc{Top++2.0}~\cite{Kidonakis:2011wy,Kidonakis:2010tc,Kidonakis:2010ux}.

Events containing $W$ or $Z$ bosons with associated jets, including jets from the hadronisation of
$b$- and $c$-quarks, were simulated using the \SHERPA v2.2.1 generator.
Matrix elements were calculated for up to two additional partons at next-to-leading order (NLO) and four partons at leading order (LO) using the {\textsc{Comix}\xspace}~\cite{Gleisberg:2008fv} and {\textsc{Open Loops}\xspace}~\cite{Cascioli:2011va} matrix element generators and
merged with the \SHERPA PS~\cite{Schumann:2007mg}  using the ME+PS@NLO prescription~\cite{Hoeche:2012yf}. The
NNPDF30NNLO~\cite{Ball:2012cx} PDF set was used in conjunction with the dedicated PS tune developed by the \SHERPA authors.

Diboson and triboson processes were also simulated using the \SHERPA generator using the NNPDF30NNLO PDF set in conjunction with a dedicated PS tune developed by the \SHERPA authors.
Matrix elements for these samples were calculated for up to one (diboson processes)  or  zero  (triboson processes) additional partons  at  NLO  and  up  to  three (diboson processes) or two
(triboson processes) additional partons at  LO.
Additional contributions to the SM backgrounds in the signal regions arise from the production of \ttbar pairs
in association with $W$/$Z$/$h$ bosons and possibly additional jets.  These processes were modelled by event samples generated at NLO
using the {\textsc{MadGraph5\_aMC\@ NLO}\xspace}~\cite{Alwall:2014hca} v2.2.3 generator and showered with the \PYTHIA v8.186 PS.

\begin{table*}
\caption{Simulated signal and background event samples: the corresponding generator, parton shower, cross-section normalisation, PDF set and underlying-event tune are shown.\vspace{-0.3cm}}
\label{tab:MC}
\begin{center}
\resizebox{\textwidth}{!}
{\small
\begin{tabular}{l l c c c c}
\toprule
Physics process    & Generator & Parton shower & Cross-section & PDF set & Tune \\
&	      & 	      & normalisation & 	&      \\
\midrule
Dark-matter signals    	   & \MADGRAPH 2.3.3~\cite{Alwall:2014hca} & \PYTHIA 8.212~\cite{Sjostrand:2007gs} & NLO~\cite{Mattelaer:2015haa,Backovic:2015soa} & NNPDF23LO~\cite{Lai:2010vv} & A14~\cite{pub-2014-021} \\
\midrule
$W(\rightarrow \ell\nu)$ + jets              & \SHERPA 2.2.1~\cite{Gleisberg:2008ta}    & \SHERPA 2.2.1     & NNLO~\cite{Catani:2009sm}  & NNPDF30NNLO~\cite{Lai:2010vv}    & \SHERPA default \\
$Z/\gamma^{*}(\rightarrow \ell \ell)$ + jets & \SHERPA 2.2.1   & \SHERPA 2.2.1       & NNLO~\cite{Catani:2009sm}  & NNPDF30NNLO    & \SHERPA default\\ \midrule

$t\bar{t}$              & {\sc powheg-box} v2~\cite{Alioli:2010xd} & \PYTHIA6.428~\cite{Sjostrand:2006za}  & NNLO+NNLL~\cite{Czakon:2013goa,Czakon:2012pz,Czakon:2012zr,Baernreuther:2012ws,Cacciari:2011hy,Czakon:2011xx}  & NLO CT10~\cite{Lai:2010vv}  &\sc{Perugia2012}~\cite{Skands:2010ak}\\

Single-top              &&&&\\
($t$-channel)           & {\sc powheg-box} v1 & \PYTHIA6.428 & NNLO+NNLL~\cite{Kidonakis:2011wy}  & NLO CT104f  & \sc{Perugia2012}\\
Single-top              &&&&\\
($s$- and $Wt$-channel) & {\sc powheg-box} v2 & \PYTHIA6.428  & NNLO+NNLL~\cite{Kidonakis:2010ux,Kidonakis:2010tc}  & NLO CT10   & \sc{Perugia2012}\\ \midrule

$t\bar{t}+W/Z/\gamma^{*}/h$ & \AMCATNLO 2.2.3 (NLO)  & \PYTHIA 8.186	& NLO~\cite{Alwall:2014hca}   & NNPDF30NLO  & A14    \\
Diboson    	   & \SHERPA 2.2.1~\cite{Gleisberg:2008ta}        & \SHERPA 2.2.1	& NLO	 & NNPDF30NNLO	     & \SHERPA default \\
\midrule
$h+W/Z$	           & \AMCATNLO 2.2.3  (NLO)     & \PYTHIA 8.186   & NLO~\cite{Dittmaier:2012vm}   & NNPDF30NLO      & A14  \\
$t\bar{t}+WW/t\bar{t}$  & \AMCATNLO 2.2.3  (LO)     & \PYTHIA 8.186   & NLO~\cite{Alwall:2014hca}	 & NNPDF23LO	  & A14  \\
$t+Z/WZ/t\bar{t}$    & \AMCATNLO 2.2.3  (LO)     & \PYTHIA 8.186   & LO                   	    & NNPDF23LO     & A14  \\
Triboson	   & \SHERPA 2.2.1        & \SHERPA 2.2.1    & NLO            & NNPDF30NNLO	     & \SHERPA default \\
\bottomrule
\end{tabular}
}
\end{center}
\end{table*}

In all MC samples, except those produced by \SHERPA, the {\sc EvtGen}~v1.2.0 program~\cite{EvtGen} was used to model the properties of the bottom and charm hadron decays.
All \PYTHIA v6.428 samples used the PERUGIA2012~\cite{Skands:2010ak} tune for the underlying event, while \PYTHIA
v8.186 and Herwig++ showering were run with the A14 and UEEE5~\cite{Gieseke:2012ft} underlying-event tunes, respectively.
To simulate the effects of additional \emph{pp} collisions in the same and nearby bunch crossings, additional interactions were generated using the soft QCD processes of \PYTHIA 8.186
with the A2 tune~\cite{ATL-PHYS-PUB-2012-003} and the MSTW2008LO PDF~\cite{Martin:2009iq}, and overlaid onto each simulated hard-scatter event.

Alternative samples are employed to derive systematic uncertainties associated with the specific configuration of the MC generators used for the nominal SM background
samples, as detailed in Sect.~\ref{sec:systematics}. They include variations of the renormalisation and factorisation scales, the CKKW-L matching~\cite{Lonnblad:2011xx}
scale, as well as different PDF sets and hadronisation models.

The event generation for the dark-matter signal samples followed the prescriptions in Ref.~\cite{Abercrombie:2015wmb}.
Events were generated from leading-order (LO) matrix elements using
the \MADGRAPH generator v2.3.3 interfaced to \PYTHIA v8.212 with the {\sc
A14} tune for the modelling of the top-quark decay chain (when applicable),
parton showering, hadronisation and the description of the underlying
event. The renormalisation and factorisation scale choice adopted is the default \MADGRAPH
dynamical scale as documented in Ref.~\cite{Hirschi:2015iia}.
For the \bbbar+$\phi/a$ and \ttbar+$\phi/a$ models the events were
generated with up to one additional parton, while for the \bFDM\
models the events were generated with up to two additional partons.
The \ttbar+$\phi/a$ and \bFDM\ samples were generated in the 5-flavour
scheme, while the \bbbar+$\phi/a$ samples were generated in the
4-flavour scheme.
Following Ref.~\cite{Abercrombie:2015wmb}, the minimum \pt\ requirement for \bjets\ in the final state
in \MADGRAPH   was set to $30\;\GeV$ for the \bbbar+$\phi/a$ model,
in order to increase the number of events in the relevant phase space for the
analysis.
This requirement does not affect the MC signal sample passing the event selection.
The PDF set NNPDF23LO was used, adopting $\alphas = 0.130$ and either the 5-flavour or the 4-flavour scheme
consistently with the choice made for generating the events.
The jet--parton matching was realised following the CKKW-L prescription.
For the \ttbar+$\phi/a$ model the matching scale was set to one quarter
of the mass of the particle mediating the interaction between the SM
and DM sectors. For the \bbbar+$\phi/a$ and \bFDM\ models
the matching scale was set to 30 GeV.
The coupling $g$ between the colour-neutral mediator for
the \ttbar+$\phi/a$ and \bbbar+$\phi/a$ models and both the SM and the
dark sector
was assumed to be one, which implies pure Yukawa-type couplings between the mediator
and the SM quarks. This choice impacts the mediator width and cross-section
calculation for these models, but it was shown to have
no significant impact on the kinematic
properties~\cite{Abercrombie:2015wmb}.

For the \ttbar+$\phi/a$  and \bbbar+$\phi/a$ models the production
cross-section was computed at NLO accuracy in the strong coupling
constant $\alphas$ using the \AMCATNLO generator with the
NNPDF30NLO PDF set using $\alphas = 0.118$. For this procedure a
dynamical scale equal to $P_\mathrm{T}/2$ was adopted, with $P_\mathrm{T}$ being
the scalar sum of the transverse momenta of all final-state particles. The
flavour scheme adopted is consistent with that used for event
generation.
For the mass range in which this analysis is sensitive, the NLO value
of the cross-sections for the \ttbar+$\phi/a$ model is about 25\%
larger than the corresponding LO value~\cite{Mattelaer:2015haa,Backovic:2015soa}.
For the \bbbar+$\phi/a$ samples the NLO value of the cross-section is
between 56\% and 75\% of the corresponding LO value.
This is driven by the \MADGRAPH\ minimum \bjet\ \pt\ requirement due to the strong dependence
of the NLO cross-section on this parameter.
For the \bFDM\ signal models, the cross-section was computed at LO
accuracy using the \AMCATNLO generator and the same flavour scheme used for
the event generation.

\section{Event selection}
\label{sec:selection}

Five signal regions (SR) are defined and optimised to detect
dark-matter production via spin-0 mediators. Two signal regions, \sroneb\ and \srtwob,
are optimised for models in which dark matter is produced in conjunction
with one or two $b$-quarks, respectively.
Specifically, \sroneb\ is designed
to optimally select candidate signal events of the colour-charged
scalar mediator models (bFDM) introduced in Sect.~\ref{sec:intro}.
\srtwob\ focuses instead on scalar and pseudoscalar colour-neutral mediators
and was specifically optimised for low mediator masses (below $200\;\GeV$).
These SRs require events with no leptons and low jet multiplicity.
\srtlow ,
\srthigh\ and
\srttwol\ are optimised to detect events in which
DM is produced in association with a \ttbar\ pair, which either decays
fully hadronically (\srtlow\ and \srthigh) or dileptonically (\srttwol). The \srtlow\ and \srthigh\ SRs are optimised for
low ($< 100\;\GeV$) and high (between $100$ and $350\;\GeV$) mediator mass assumptions, respectively, and are assigned fully hadronic events
with high jet multiplicity. The regions  \srtlow\  and \srthigh\
overlap in terms of their selection criteria.
The region \srttwol\ focuses on  mediator masses below $100\;\GeV$
and contains events with two leptons in the final state.

\subsection{Signatures with $b$-quarks and \met}

Events assigned to \sroneb\ and \srtwob\ are required to pass the missing transverse momentum
trigger and to have at least one jet ($\mathcal{N}_j$).
A minimum azimuthal angle between the directions of the missing transverse momentum and any of the jets in the event ($\dphimin$)
is required, in order to reduce the contamination by multi-jet
events where fake \met\ arises from jet energy mismeasurements or semileptonic decays of hadrons inside jets.
Events with at least one baseline muon or electron ($\mathcal{N}^{\mathrm B}_{\ell}$) are
discarded to reject leptonic decays of $W$ and $Z$ bosons.
The dominant background processes for the events passing these requirements are \ttbar\ and \Zjets\ processes.

Events with at least one tight \btagged\ jet
($\mathcal{N}^{\mathrm T}_{b}$) and which pass the kinematic
requirements specified in Table~\ref{t:srbselections} are assigned to \sroneb.
The high-\met\ selection required is essential to discriminate
the signal from the background in this SR.
An upper limit on the scalar sum of the transverse momenta of the baseline jets in the events excluding the
leading and subleading jets (\httree~\cite{Sbottom}) is used in this
SR to reduce the contributions from top-quark pair-production processes.

Events assigned to \srtwob\ have instead at least two tight
\btagged\ jets. When the \btagged\ jet multiplicity is different from two, the
\btagged\ jets are sorted in descending order according to their $b$-tagging probability.
For this SR,
a requirement of low jet multiplicity
was found to be more effective in reducing the \ttbar\ background.
The jet multiplicity of candidate signal events is required to not exceed three, and
the transverse momentum of the third jet in the event must not exceed
$60\;\GeV$. For the same purpose, the ratio of the transverse momentum
of the leading jet to \Ht,
the scalar sum of the transverse momenta of all jets in the events,
($\htratio = \pt(j_1) / \Ht$) is required to be larger than 75\%.

\begin{table}[t]
\caption{Summary of the kinematic and topology-dependent selections for signal regions \sroneb\ and
\srtwob .}
\centering
\begin{tabular}{lll}
\toprule
Observable & \sroneb & \srtwob \\
\midrule
Trigger                                 &  \multicolumn{2}{c}{\met} \\
$\mathcal{N}_{j}$   & $\geq2$ & $2$ or $3$ \\
$\mathcal{N}^{\mathrm T}_{b}$   & $\geq1$ & $\geq2$ \\
$\mathcal{N}^{\mathrm B}_{\ell}$ & \multicolumn{2}{c}{$0$}\\
\midrule
\met\ [\GeV]         &$>650$   & $>180$  \\
$\pt(bj_1)$ [\GeV]    &$>160$  &$>150$   \\
$\pt(j_1)$ [\GeV]     &$>160$  &$>150$  \\
$\pt(j_2)$ [\GeV]     &$>160$  &$>20$  \\
$\pt(j_3)$  [\GeV]     & -      &$<60$  \\
\midrule
\httree\ [\GeV]     & $< 100 $ & -        \\
\htratio      & - &$>0.75$          \\
\xone\ [rad]        & - & $< 0$             \\
\yone\ [rad]        & - &$<0.5$         \\
\midrule
\multicolumn{3}{l}{Multi-jet rejection specific} \\
\dphimin\ [rad]     & $>0.6$ & $>0.4$ \\
\bottomrule
\end{tabular}
\label{t:srbselections}
\end{table}

The azimuthal separations between the \btagged\ jets (\dphibb) and the
\dphimin\ are exploited to enhance the separation between
the signal and the irreducible background in this channel
($Z(\nu\bar{\nu})$+$b\bar{b}$), as the latter
is characterised by small \dphibb\ values when the \bjets\ originate
from the gluon-splitting process. Linear
combinations of these two variables are used to define the
selection criteria in Table~\ref{t:srbselections}:
\begin{align*}
\xone &= \dphimin - \dphibb , \\
\yone &= \vert\dphimin + \dphibb - \pi \vert.
\end{align*}
An additional handle to discriminate
between the $b\bar b+\phi$ and $b\bar b+a$ signal models and the background
is the spin of the particle decaying into invisible decay products.
It was shown in Ref.~\cite{Haisch:2016gry} that it is possible to discriminate
between such scalar, pseudoscalar and vector particles by exploiting
information about the production angle of the visible particles with respect to the proton beam axis.
A convenient variable to exploit
this feature, proposed in Ref.~\cite{Barr:2005dz} relies on the pseudorapidity difference between the two \btagged\ jets\ ($\Delta\eta_{bb}$):
\begin{equation*}
\cthbb = \left|\tanh{\left(\frac{\Delta\eta_{bb}}{2}\right)}\right|.
\end{equation*}
The variable \cthbb , evaluated in the laboratory frame,
is the key observable used in \srtwob\ to discriminate the signal
from the background.
The distribution of \cthbb\ is approximately flat for
\bjets\ produced in association with scalar or vector particles with masses
below $100\;\GeV$, while it exhibits a pronounced
enhancement at values near one for pseudoscalar particles
in the same mass range.
In order to further enhance the sensitivity to the signal,
the signal region \srtwob\  is divided into four independent bins in \cthbb:
\srtwob-bin1 $(0,0.25)$, \srtwob-bin2 $(0.25,0.5)$, \srtwob-bin3 $(0.5,0.75)$, \srtwob-bin4 $(0.75,1.0)$,
which are statistically combined in the final result.

\subsection{Signatures with top quarks and \met}

\begin{table}[p]
\caption{Summary of the kinematic and topology-dependent selections for signal regions \srtlow , \srthigh\ and \srttwol .}
\centering
\begin{tabular}{lccc}
\toprule
Observable  & \srtlow & \srthigh & \srttwol\\[0.3ex]
\midrule
Trigger             &   \multicolumn{2}{c}{\met}  & $ 2\ell$ \\[0.3ex]
$\mathcal{N}_{j}$   & \multicolumn{2}{c}{$\geq4$} & $\geq1$\\[0.3ex]
$\mathcal{N}^{\mathrm M}_{b}$  & \multicolumn{2}{c}{$\geq2$} & $\geq1$\\[0.3ex]
$\mathcal{N}^{\mathrm B}_{\ell}$  & \multicolumn{2}{c}{$0$} & - \\[0.3ex]
$ \mathcal{N}^{\mathrm M}_{\ell}$ & \multicolumn{2}{c}{-} & $2$ OS\\[0.3ex]
$ \mathcal{N}_{\tau}$ &  \multicolumn{2}{c}{$0$} & - \\[0.3ex]
\midrule
\met\ [\GeV]         &\multicolumn{2}{c}{$>300$} & -    \\[0.3ex]
$\pt(bj_1)$ [\GeV] &  \multicolumn{2}{c}{$>20$} &$>30$    \\[0.3ex]
$\pt(j_1, j_2)$ [\GeV]  & \multicolumn{2}{c}{$>80, 80$} &$>30$   \\[0.3ex]
$\pt(j_3, j_4)$  [\GeV] & \multicolumn{2}{c}{$>40, 40$} &- \\[0.3ex]
\midrule
$\pt(\ell_1, \ell_2)$  [\GeV]      & \multicolumn{2}{c}{-} &$>25, 20$\\[0.3ex]
$\mll$ [\GeV] & \multicolumn{2}{c}{-} & $  > 20$ \\[0.3ex]
$\vert\mll^{\mathrm{SF}} \hskip-3pt- \hskip-2pt m_\Zboson\vert$  [\GeV]& \multicolumn{2}{c}{-} & $ > 20$ \\[0.3ex]
\midrule
\remass{1,2}{0.8} [\GeV]  & $>80,80$ & -    & - \\[0.3ex]
\remass{1,2}{1.2} [\GeV]  & - & $>140,80$   & - \\[0.3ex]
\mtbmin\ [\GeV]           & $>150$ & $>200$& - \\[0.3ex]
\mtbmax\ [\GeV]           & $>250$ & -     & - \\[0.3ex]
\drbb\    & $>1.5$  &  $>1.5$      & - \\[0.3ex]
\metsig\  [$\sqrt{\GeV}$] & - & $>12$ & - \\
\midrule
\dphib\ [rad]          & \multicolumn{2}{c}{-} & $< 0.8$ \\[0.3ex]
\minbl\ [\GeV ]       & \multicolumn{2}{c}{-} & $<170$\\[0.3ex]
\cem\ [\GeV]     & \multicolumn{2}{c}{-}  & $>170$\\[0.3ex]
\mttwoll\ [\GeV]       & \multicolumn{2}{c}{-} & $>100$\\[0.3ex]
\midrule
\multicolumn{4}{l}{Multi-jet rejection specific} \\
\dphimin\ [rad]     &  \multicolumn{2}{c}{$>0.4$} & -\\
\mettrack [\GeV]   &  \multicolumn{2}{c}{$> 30$}  & -\\
$\Delta\phi(\ptmiss,\ptmisstrack)$  [rad] & \multicolumn{2}{c}{$< \pi/3$} & -\\
\bottomrule
\end{tabular}
\label{t:srtselections}
\end{table}

Events assigned to \srtlow\ and \srthigh\ are required to contain at
least four jets.
At least two
jets in every event must be \btagged\ at the medium working-point ($\mathcal{N}^{\mathrm M}_{b}$).
Events containing baseline electrons and muons are discarded.
Furthermore, events with a $\tau$-candidate are also rejected ($\mathcal{N}_\tau = 0$).
The $\tau$-candidate is defined as a jet with less than
four associated tracks which has not passed the medium $b$-tagging requirement
and which has a $\phi$ separation from the \ptmiss\ of no more than $\pi/5$ radians.
Events are required to pass the missing transverse momentum
trigger and to satisfy $\met > 300\; \GeV$.
Also in this SRs, a minimum $\dphimin$ requirement
is applied in order to
reject events with \met\
arising from mismeasurements and
semileptonic decays of hadrons inside jets.
Further rejection of such events is achieved by
additional requirements on the missing transverse momentum computed using only the information from the tracking system
(\ptmisstrack, with magnitude \mettrack) and its angle with respect to the \ptmiss\
($\Delta\phi(\ptmiss,\ptmisstrack)$).
The dominant backgrounds for these signal regions are top-quark pair production, $Z$+jets, and the production of a
$Z$ boson in association with \ttbar.
Four main
observables
are
exploited to discriminate DM signal events from the SM background processes: \mtbmin,
\mtbmax, \metsig, and \drbb.
The variables \mtbmin\ and \mtbmax\ are defined as the transverse mass%
\footnote{The transverse mass of two particles $a$ and $b$ is defined as
$m_{\mathrm T}(a,b) = \sqrt{(E_{\mathrm T,a}+E_{\mathrm T,b})^2 - (\vec{p}_{\mathrm T,a}+\vec{p}_{\mathrm T,b})^2}$}
of the \ptmiss\ vector and \btagged\ jet
with the smallest and largest angular distance%
\footnote{The angular separation between two particles $a$, $b$ used in all quantities described in this section is
defined as $\Delta R_{ab} = \sqrt{(\Delta\phi_{ab})^2+(\Delta\eta_{ab})^2}$. }
from it,
respectively.
The \mtbmin\ variable is designed to be bounded from above by the top-quark mass for semileptonic \ttbar\ decays,
because the closest \btagged\ jet to the \ptmiss\ vector usually belongs to the leg of the decay where the \Wboson\ boson decays into leptons.
The variable \mtbmax\ recovers the discriminating power in the case of wrong pairing.
The \metsig\ variable is defined as the ratio of the \met to the square-root of the scalar sum of the
transverse momenta of all jets in the events (\Ht)  to discriminate the
high-mediator-mass signal models in \srthigh\ from the SM background.
Finally, the angular distance between the two \btagged\ jets in the
event (\drbb)
is exploited to suppress $Z(\nu\nu)$+$b\bar b$ events where the two $b$-quarks
arise from gluon-splitting and are characterised by a small angular separation.

The \srtlow\ selection is optimised for low-mass spin-0
mediators ($m(\phi/a) < 100\;\GeV$).
Requirements on the two leading reclustered jet masses with
radius $0.8$ (\remass{1}{0.8}, \remass{2}{0.8})
exploit the presence
of boosted hadronic decays of \Wboson\ bosons from top quarks in the event.
The requirements applied in \srtlow\ are such that both reclustered jets are
compatible with a $W$-boson candidate.
The \srthigh\ signal region is optimised instead for high-mass spin-0
mediators ($100\;\GeV < m(\phi/a) < 350\;\GeV$).
Requirements on the two leading reclustered jet masses with
radius $1.2$ (\remass{1}{1.2}, \remass{2}{1.2}) are used to
exploit
the more boosted topology of
these signal events compared to the backgrounds.
The requirements applied in \srthigh\ are such that the leading
large-radius jet is compatible with a top-quark candidate and the
subleading large-radius jet is compatible with a $W$-boson candidate.
The specific requirements for each discriminating
observable in \srtlow\ and \srthigh\ are summarised in Table~\ref{t:srtselections}.

Finally, events assigned to \srttwol\ are required to have exactly two
opposite-sign leptons ($\mathcal{N}^{\mathrm M}_{\ell} = 2$ OS), electrons or muons, either same- or different-flavour, with an
invariant mass (regardless of the flavours of the leptons in the
pair), $m_{\ell\ell}$, being larger than $20\;\GeV$.
In addition, for
same-flavour lepton pairs, events with
$m_{\ell\ell}$
within $20\; \GeV$ of the \Zboson-boson mass are vetoed.
Furthermore, candidate signal events are required to have
at least
one medium \btagged\ jet.
Events are
required to pass the two-lepton triggers and the leading and subleading
lepton transverse momenta in the event are required to be at least $25$ and $20
\;\GeV$, respectively, which also guarantees that the plateau of efficiency of the triggers
is reached.
The main reducible backgrounds for this analysis are dileptonic \ttbar\ decays, \Zjets and dibosons.
The main handle for the rejection of these backgrounds is the lepton-based "stransverse mass",
\mttwoll~\cite{Stop2LRun1,Barr:2003rg,Lester:1999tx},
which is a kinematic variable with an endpoint at the $W$-boson mass for events
containing two \Wboson\ bosons decaying into leptons.
In this selection it is used in
linear combination with the \met, in order to maximise the
discrimination power of the two variables~\cite{Haisch:2016gry}:
\begin{equation*}
\cem = \mttwoll + 0.2 \cdot \met.
\end{equation*}
Further requirements are placed on $\dphib$~\cite{Stop2LRun1}, the azimuthal angular distance
between \ptmiss\ and the vector sum of \ptmiss\ and the transverse momentum of the leptons,
and on $\minbl$, which is the smallest invariant
mass computed between the \btagged\ jet and each of the two leptons in the event.
Both variables are used to further reject residual contamination from reducible backgrounds for this selection.
The variable \dphib, can be interpreted as the azimuthal angular difference between the \ptmiss
and the opposite of the vector sum of all the transverse hadronic activity in the event.
The requirement on this variable reject $Z(\ell^+\ell^-)$+jets events where the \met arises from
jet mismeasurements, while retaining a large fraction of the signal.
In events with two top quarks decaying dileptonically such as in the signal topology, at least one of the
two mass combinations must be bounded from above by $\minbl< \sqrt{m_t^2 - m_W^2}$. This variable
helps to reject residual reducible backgrounds, while retaining 99\% of the signal.
The specific requirements for \srttwol\ are
summarised in Table~\ref{t:srtselections}.

\section{Background estimation}
\label{sec:bkg}

The SM backgrounds contributing to each of the five SRs are estimated with the aid of the MC simulation and using
control regions (CRs) constructed to enhance a particular background and to be kinematically similar but
orthogonal to the SRs.
The expected background is determined separately in each SR through a profile likelihood fit based on the HistFitter package~\cite{Baak:2014wma}.
The CR yields constrain the normalisation of the dominant SM background  processes.
Such normalisation factors are treated as free fit parameters and are uncorrelated between fits of different SRs.
The systematic uncertainties are included as nuisance parameters in the fit.
In the case of a "background-only" fit set-up, only the CRs are considered and the signal
contribution is neglected.
The number of background events predicted by simulation in the SRs
is normalised according to the results of the fit.
When computing exclusion limits as described in Sect.~\ref{sec:result}, the SRs are also
used to constrain the background predictions.
The non-dominant SM
backgrounds are determined purely from MC simulation, except fake or non-prompt lepton backgrounds
(arising from jets misidentified as leptons or produced in either hadron decays or photon conversions)
and the multi-jet background, both of which are estimated using a data-driven method described below.
The background estimates in the SRs are validated by extrapolating
the results of the likelihood fit in the CRs
to dedicated validation regions (VRs),
which are designed to be orthogonal to both the signal and control regions. In all CRs and VRs used in this analysis the signal contamination was found to be
negligible.

An important source of background for all 0-lepton signal regions is
$Z$ bosons decaying into neutrinos when produced in conjunction with one or
more jets emanating from heavy-flavour quarks. Production of top-quark pairs is a substantial background source
for all selections except for \sroneb, where the very high \met\ requirement rejects this background.
More specifically, top-quark pairs with at least one of the $W$ bosons decaying into leptons (where the lepton is either a non-identified electron or muon,
or a hadronically decaying $\tau$ lepton) enter \srtwob , \srtlow\ and \srthigh , while events with both $W$-bosons decaying into leptons enter \srttwol .
Events from \ttZ\ production, when the
$Z$ boson decays into neutrinos, are an irreducible background for the three SRs targeting dark matter produced in association with top quarks.

The normalisation factor for the background
arising from $Z\rightarrow\nu\bar{\nu}$ events is estimated from data
in CRs with two tight same-flavour opposite-sign (SFOS)
leptons ($\ell = (e, \mu)$) and an invariant mass compatible with the
$Z$-boson mass. For these CRs, labelled in the following
as CRZ\apptlow, CRZ\appthigh, CRZ\apponeb\ and CRZ\apptwob, the \pt of the
leptons is added vectorially to the
\ptmiss\ to mimic the expected
missing transverse momentum spectrum of $Z\rightarrow\nu\bar{\nu}$
events, and is denoted in the following by \metprime.
Observables that make use of \met\ in their definition are recalculated
for these regions by using \metprime\ instead. These variables are
\xoneprime, \yoneprime, \dphiminprime, \mtbminprime, \mtbmaxprime\ and \metsigprime.

Single tight-lepton CRs, denoted by CRT\apptwob, CRT\apptlow\ and CRT\appthigh , are used to estimate the background from top-quark pairs in \srtwob, \srtlow\ and \srthigh .
The transverse mass\footnote{%
The transverse mass in this case is calculated by neglecting the lepton masses.%
} (\mtlep) of the lepton and the \ptmiss , and the angular distance between the lepton and the
\btagged\ jet closest to it (\drblep) are used to enhance the purity of
top-quark events.
In CRT\apptlow\ and CRT\appthigh\ the lepton is treated as
a jet, in order to better mimic the type of background events that contaminate the corresponding
SR. The dileptonic top background,
which contaminates \srttwol , is instead estimated in
a two-medium-leptons CR composed of events that fail the
\cem\ requirement (CRT\appttwol).

Finally, \ttV\ events, and in particular \ttZ\ events where the $Z$ boson decays into neutrinos,
represent the irreducible background
for the three SRs targeting dark matter produced in association with top quarks.
This background is estimated from data using two CRs.
To estimate the normalisation factor for the \ttZ\ background in \srtlow\ and \srthigh\
a control region of \ttbar+$\gamma$ events  (\crgamma) is used.
Events with $p_{{\mathrm{T}}\gamma} > m_{\mathrm Z}$ are selected, for which the kinematic properties
resemble those of $\ttZ(\nu\nu)$. The \crgamma\ contains
events with exactly one energetic tight photon ($\mathcal{N}_\gamma = 1$) and at least one
lepton from the decay of the \ttbar\ system. This strategy substantially
increases the number of events at large missing transverse
momentum and allows \crgamma\ to better mimic the hard kinematic requirements
of \srtlow\ and \srthigh.
Furthermore, the \pt of the
photon is added vectorially to the
\ptmiss\ to mimic the expected
missing transverse momentum spectrum of $Z\rightarrow\nu\bar{\nu}$
events. The variable obtained with this procedure is referred to as \metprimeg\ in the following.

\begin{landscape}
\begin{table}
\centering
\caption{Summary of the control region selections. Only the topological requirements modified with respect to Tables~\ref{t:srbselections} and~\ref{t:srtselections} are indicated. The symbol $\mathcal{N}_{b}$ refers to either $\mathcal{N}^{\mathrm M}_{b}$ or $\mathcal{N}^{\mathrm T}_{b}$ in order to be consistent with the  SR definition for each region. }
\label{t:CRcuts}
\scalebox{0.94}{
\begin{tabular}{lcccccccccc}
\toprule
Observable & CRZ\apponeb & CRZ\apptwob & CRZ\apptlow & CRZ\appthigh & CRT\apptwob  & CRT\apptlow & CRT\appthigh & CRT\appttwol & \crgamma & \crthreel\\
\midrule
Trigger&  1$\ell$      & 1$\ell$       & \multicolumn{2}{c}{1$\ell$}  & 1$\ell$        & \multicolumn{2}{c}{\met}   & $2\ell $ & $1\gamma$ & $2\ell $\\
$\begin{array}{l}\hskip-3pt\mathcal{N}_{j} \\\hskip-3pt\mathcal{N}_{b} \end{array}$ &
$\begin{array}{c} \geq2 \\ \geq1 \end{array}$ &
$\begin{array}{c} 2\hbox{--}3 \\  \geq2 \end{array}$ &
\multicolumn{2}{c}{$\begin{array}{c} \geq4 \\ \geq2 \end{array}$} &
$\begin{array}{c}  2\hbox{--}3 \\ \geq2 \end{array}$ &
\multicolumn{2}{c}{$\begin{array}{c} \geq3 \\ \geq2 \end{array}$} &
$\begin{array}{c} \geq1 \\  \geq1 \end{array}$ &
$\begin{array}{c}  \geq4 \\ \geq2 \end{array}$ &
$\begin{array}{c} \geq3 \\ \geq2 \end{array}$ or $\begin{array}{c} \geq4 \\ =1 \end{array}$ \\
$\mathcal{N}^{\mathrm T}_{\ell}$ & \multicolumn{2}{c}{$=2$ (SFOS)} &\multicolumn{2}{c}{$=2$ (SFOS)} & $=1$& \multicolumn{2}{c}{$=1$}& -    & $=1$&-\\
$\mathcal{N}^{\mathrm M}_{\ell}$ &  - & -   &   \multicolumn{2}{c}{-}  & -   &  \multicolumn{2}{c}{-}  & $=2$ (OS) & -   &$3$ (1 SFOS)\\
$\mathcal{N}_{\tau}$ & - & - &  \multicolumn{2}{c}{0} & - & \multicolumn{2}{c}{0} & - & - & -\\
$\mathcal{N}_{\gamma}$ &  - & - &  \multicolumn{2}{c}{-} & - &   \multicolumn{2}{c}{-}  & $=1$ & - & -\\
\midrule
\met\ [\GeV]         &$<120$   & $<60$  & \multicolumn{2}{c}{$ < 50$} & $>180$ &  \multicolumn{2}{c}{$>250$} & -& - & -\\
\metprime\ [\GeV] & $> 300 $   & $> 120$&  \multicolumn{2}{c}{$>160$} & - &  \multicolumn{2}{c}{-} & -&- & $>80$\\
$\pt(\gamma)$ [\GeV]& - & - &  \multicolumn{2}{c}{-} & - &  \multicolumn{2}{c}{0} & - & $>150$& - \\
$\pt(\ell_1),\pt(\ell_2)$ [\GeV] & $>30,>25$& $>30,>25$ & \multicolumn{2}{c}{$>28,>28$} & $>30,-$ & \multicolumn{2}{c}{$>28,-$} & $>25,20$ & $>28$ & $>25,>20$ \\
Multi-jet rejection specific & \multicolumn{2}{c}{as SR} & \multicolumn{2}{c}{no} & \multicolumn{4}{c}{as SR} & no & as SR \\
\mtlep\ [\GeV]   & -   &- & \multicolumn{2}{c}{-} & $>30$   & \multicolumn{2}{c}{$[30$--$100]$} & -     & - & $>30$          \\
\drblep [rad] &  - & -  &  \multicolumn{2}{c}{-} & - & $<1.0$ & $<1.5$ & - & - & - \\
$\vert\mll \hskip-3pt - m_\Zboson\vert$\ [\GeV] & $<20$ & $<30$ & \multicolumn{2}{c}{$<5$} & - & \multicolumn{2}{c}{-} & as SR & - & $<10$\\
\midrule

\dphiminprime\ [rad] &$>0.6$ & -&\multicolumn{2}{c}{-} & -& \multicolumn{2}{c}{-}& -& -  & -   \\
\htratio & - & $>0$  &   \multicolumn{2}{c}{-} & as SR &  \multicolumn{2}{c}{-} & - & - & -\\
\xoneprime, \yoneprime\ [rad] & - & $<1, <0.5$ &   \multicolumn{2}{c}{-}& as SR &  \multicolumn{2}{c}{-} & - & - & - \\
\remass{0}{SR} [\GeV] & - & - & $>60$ & $>60$ & - & $>60$&$>140$&-&-&-\\[0.15ex]
\remass{1}{SR} [\GeV] & - & - &  \multicolumn{2}{c}{-} & - & $>60$&$>80$&-&-&-\\[0.15ex]
\mtbmin\  [\GeV] & - & - &\multicolumn{2}{c}{-} & - &\multicolumn{2}{c}{$>100$} & - & - & - \\[0.15ex]
\mtbmax\  [\GeV] & - & - & \multicolumn{2}{c}{-}& - & \multicolumn{2}{c}{-} & - & - & - \\[0.15ex]
\mtbminprime\ [\GeV] & - & - & - & $>100$ & - & \multicolumn{2}{c}{-} & - & - & - \\[0.15ex]
\mtbmaxprime\ [\GeV] & - & - & $>100$ & - & - &\multicolumn{2}{c}{-} & - & - & - \\[0.15ex]
\drbb\ & - & - &  \multicolumn{2}{c}{0} & - &  \multicolumn{2}{c}{1.5} & - & - & - \\[0.15ex]
\metsigprime\ [$\sqrt{\GeV}$] & - & - & - & $>6$ & - &  \multicolumn{2}{c}{-} & - & - & - \\[0.15ex]
\cem\  [\GeV] & - & - &  \multicolumn{2}{c}{-} & - &  \multicolumn{2}{c}{-} & $< 150$ & - & - \\[0.15ex]
\minbl\  [\GeV] & - & - &  \multicolumn{2}{c}{-} & - &  \multicolumn{2}{c}{-} & $< 170$ & - & - \\[0.15ex]
\cemprime\ [\GeV] & - & - &  \multicolumn{2}{c}{-} & - &  \multicolumn{2}{c}{-} & - & - & $>120$ \\[0.15ex]
\minblprime\ [\GeV] & - & - &  \multicolumn{2}{c}{-} & - &  \multicolumn{2}{c}{-} & - & - & $<170$ \\[0.15ex]
\bottomrule
\end{tabular}
}
\end{table}
\end{landscape}

\begin{figure}[p]
\centering
\begin{subfigure}{.48\textwidth}\centering
\includegraphics[width=.98\textwidth]{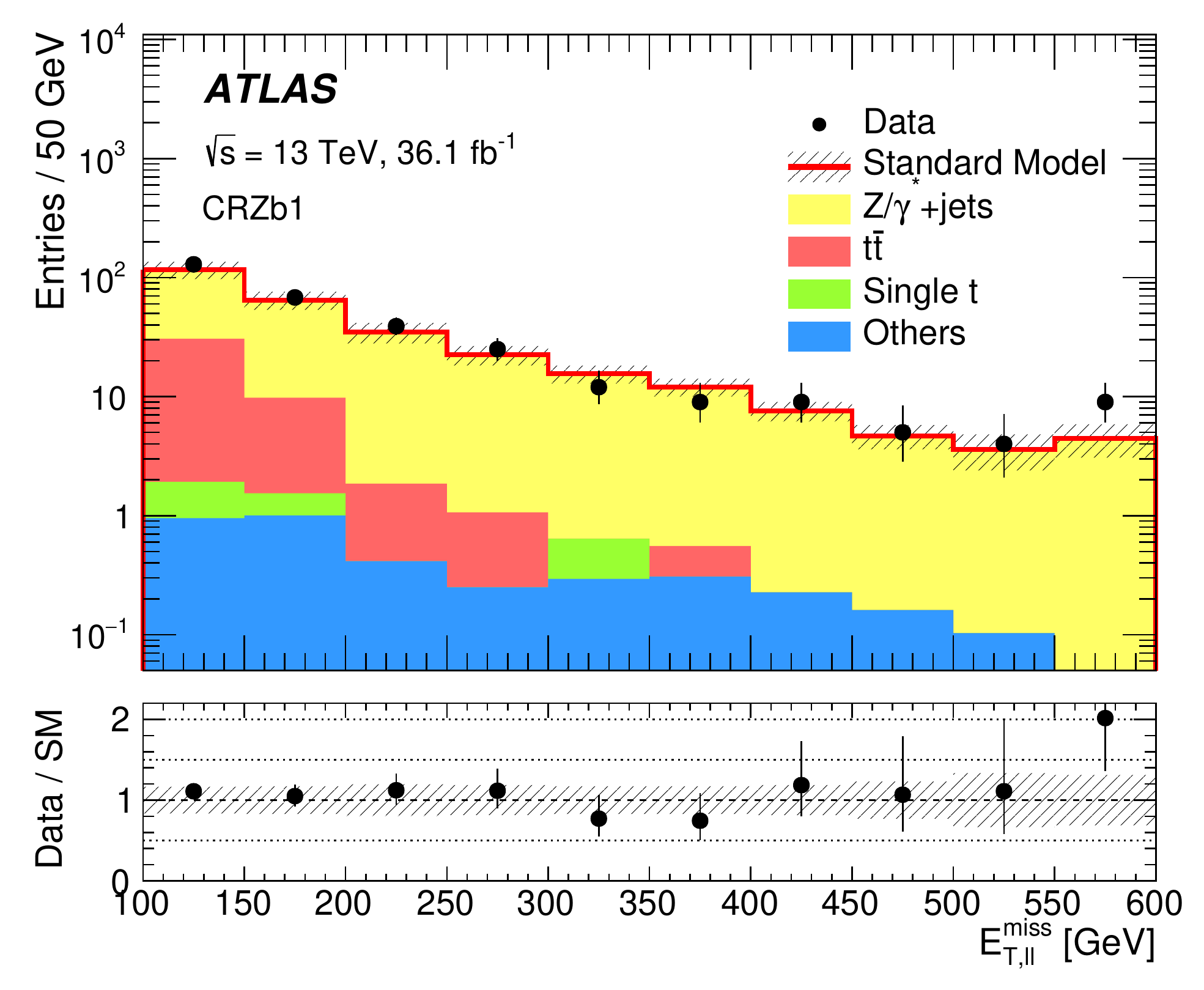}
\end{subfigure}
\begin{subfigure}{.48\textwidth}\centering
\includegraphics[width=.98\textwidth]{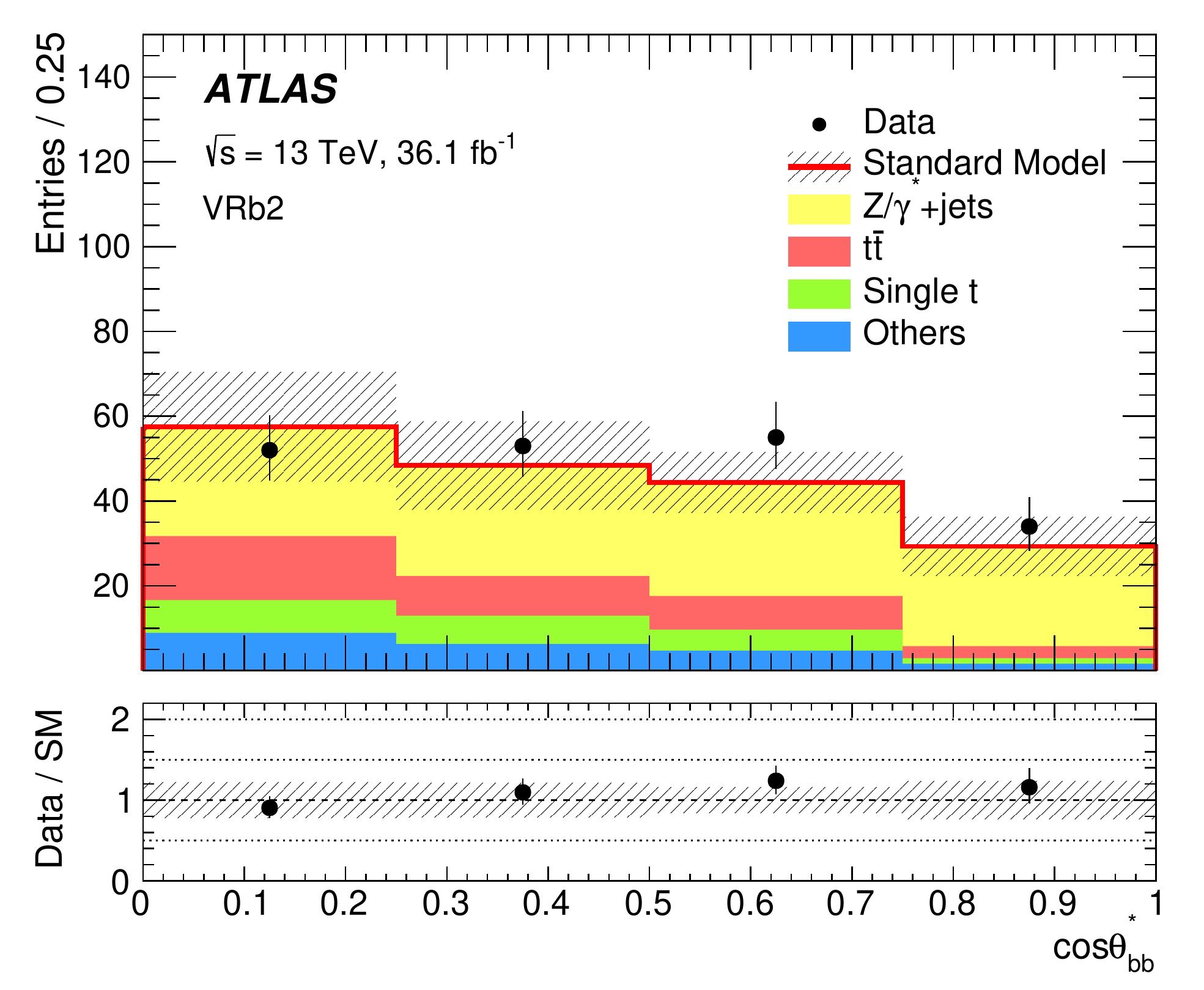}
\end{subfigure}

\begin{subfigure}{.48\textwidth}\centering
\includegraphics[width=.98\textwidth]{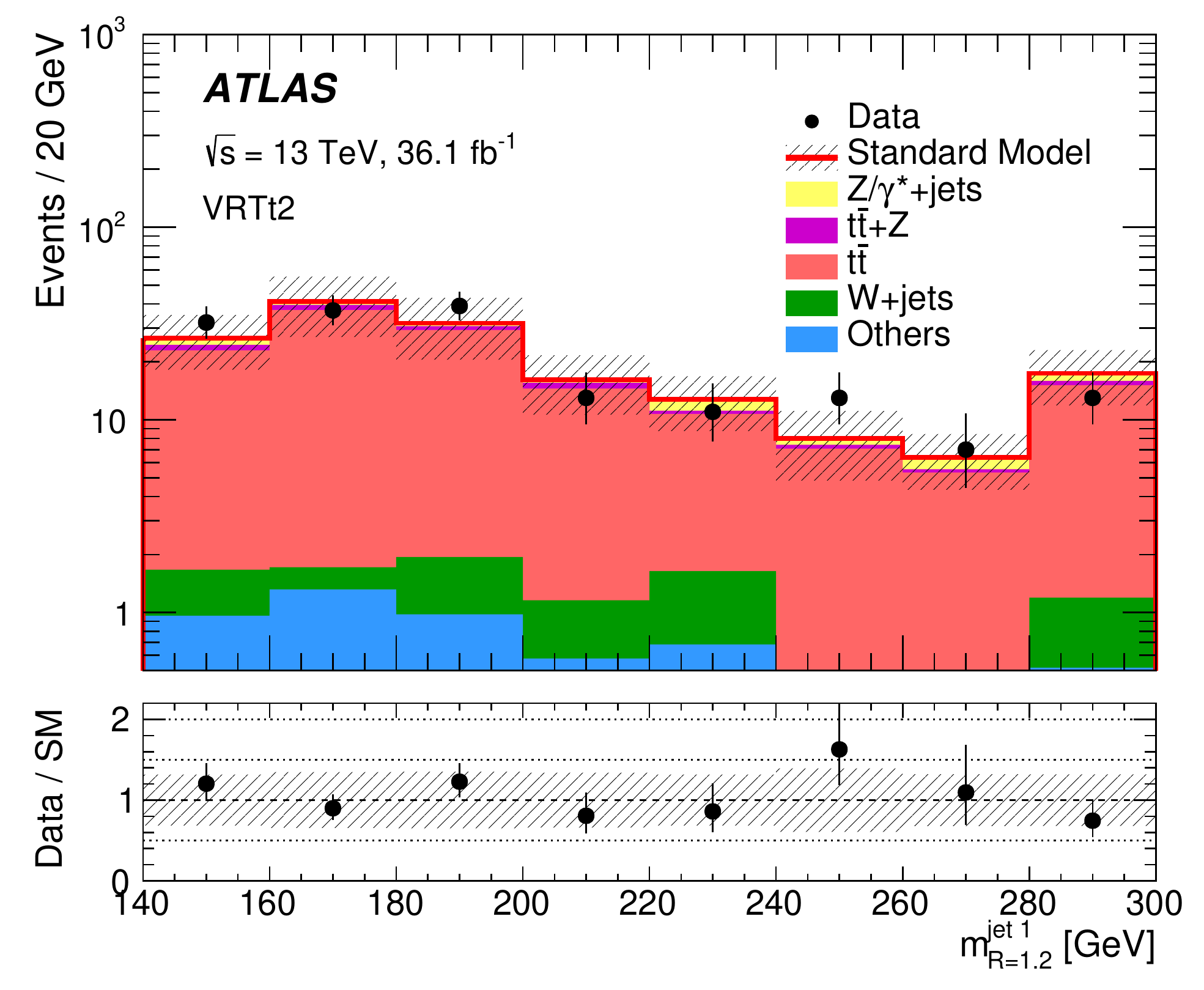}
\end{subfigure}
\begin{subfigure}{.48\textwidth}\centering
\includegraphics[width=.98\textwidth]{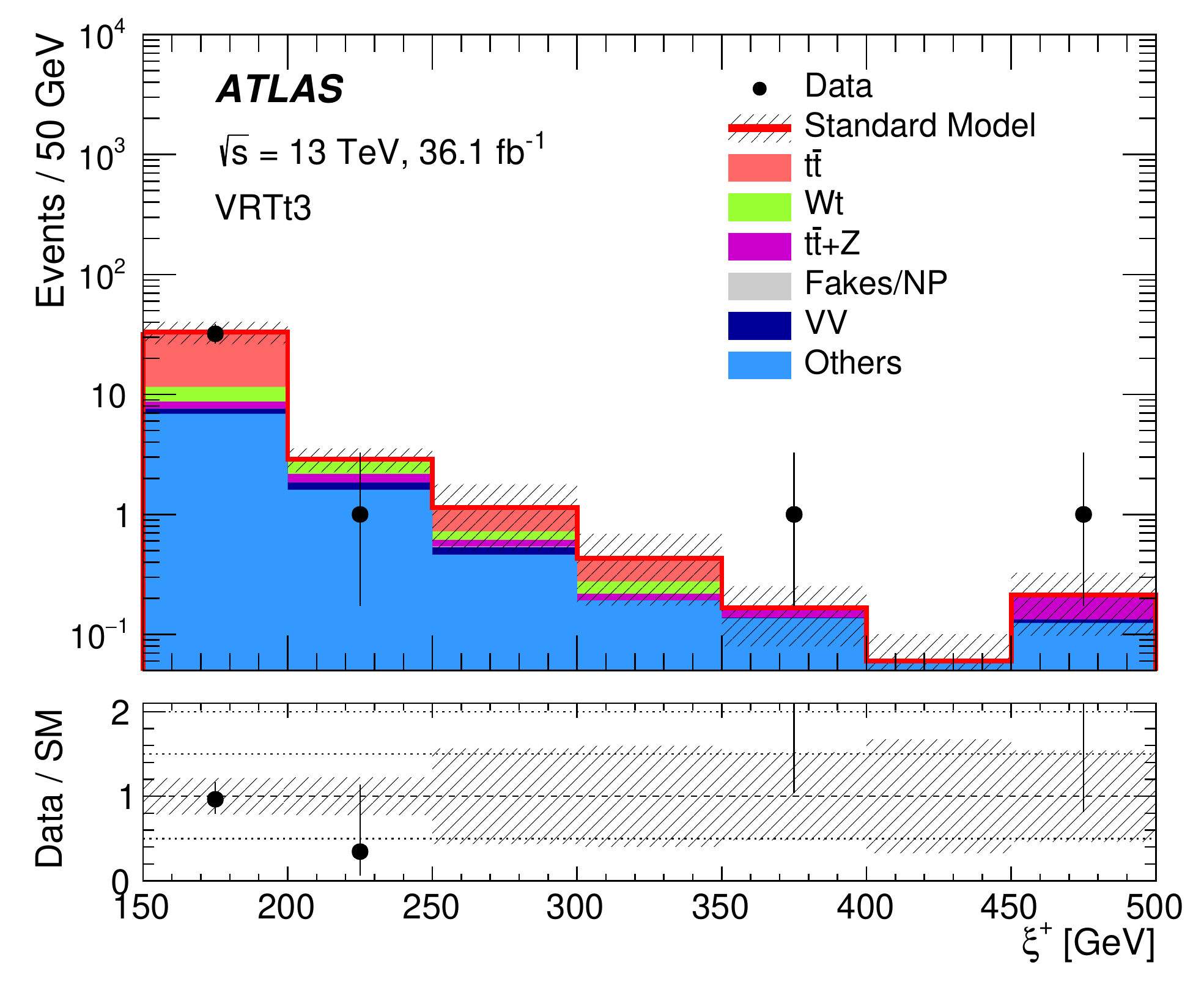}
\end{subfigure}

\begin{subfigure}{.48\textwidth}\centering
\includegraphics[width=.98\textwidth]{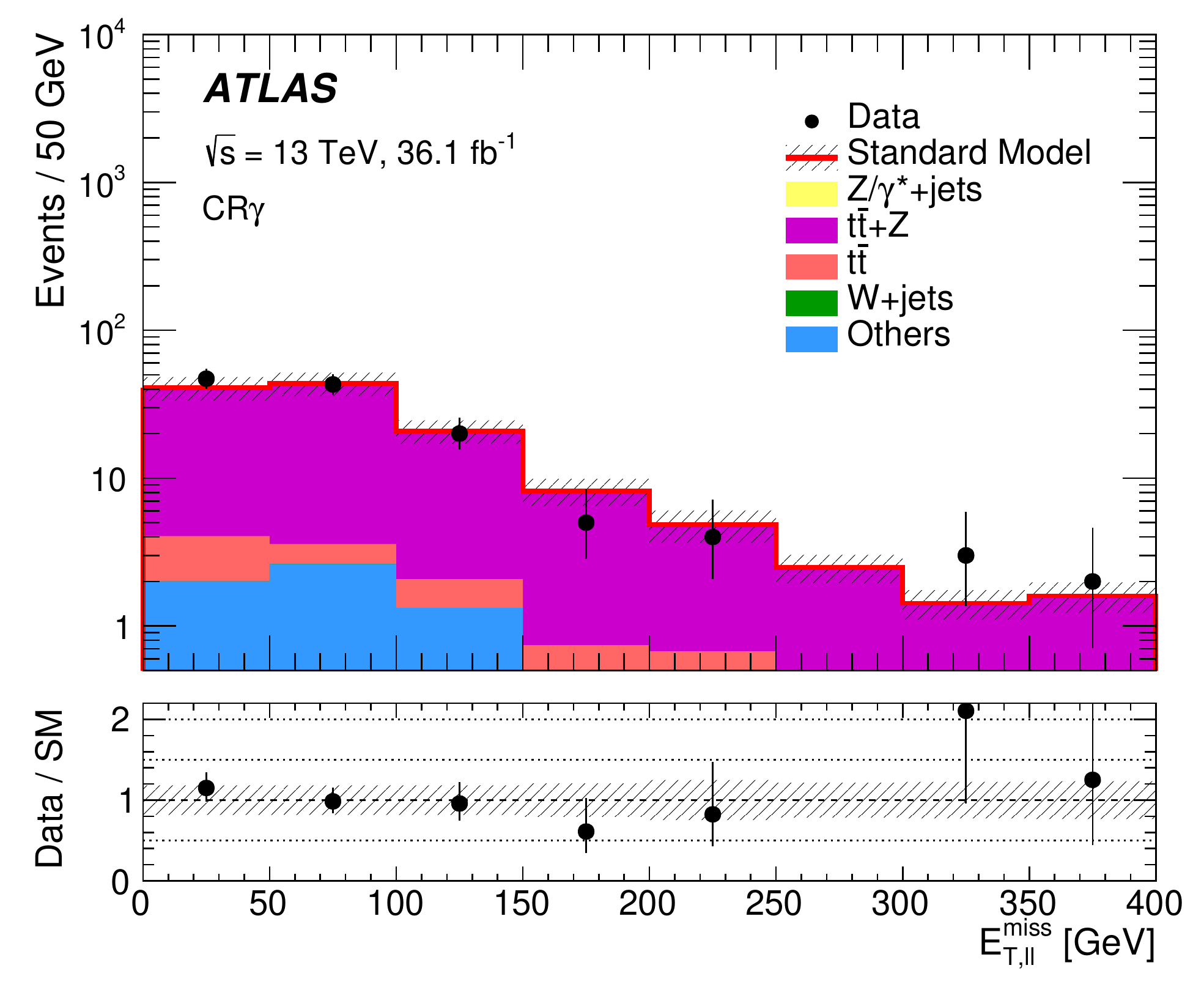}
\end{subfigure}
\begin{subfigure}{.48\textwidth}\centering
\includegraphics[width=.98\textwidth]{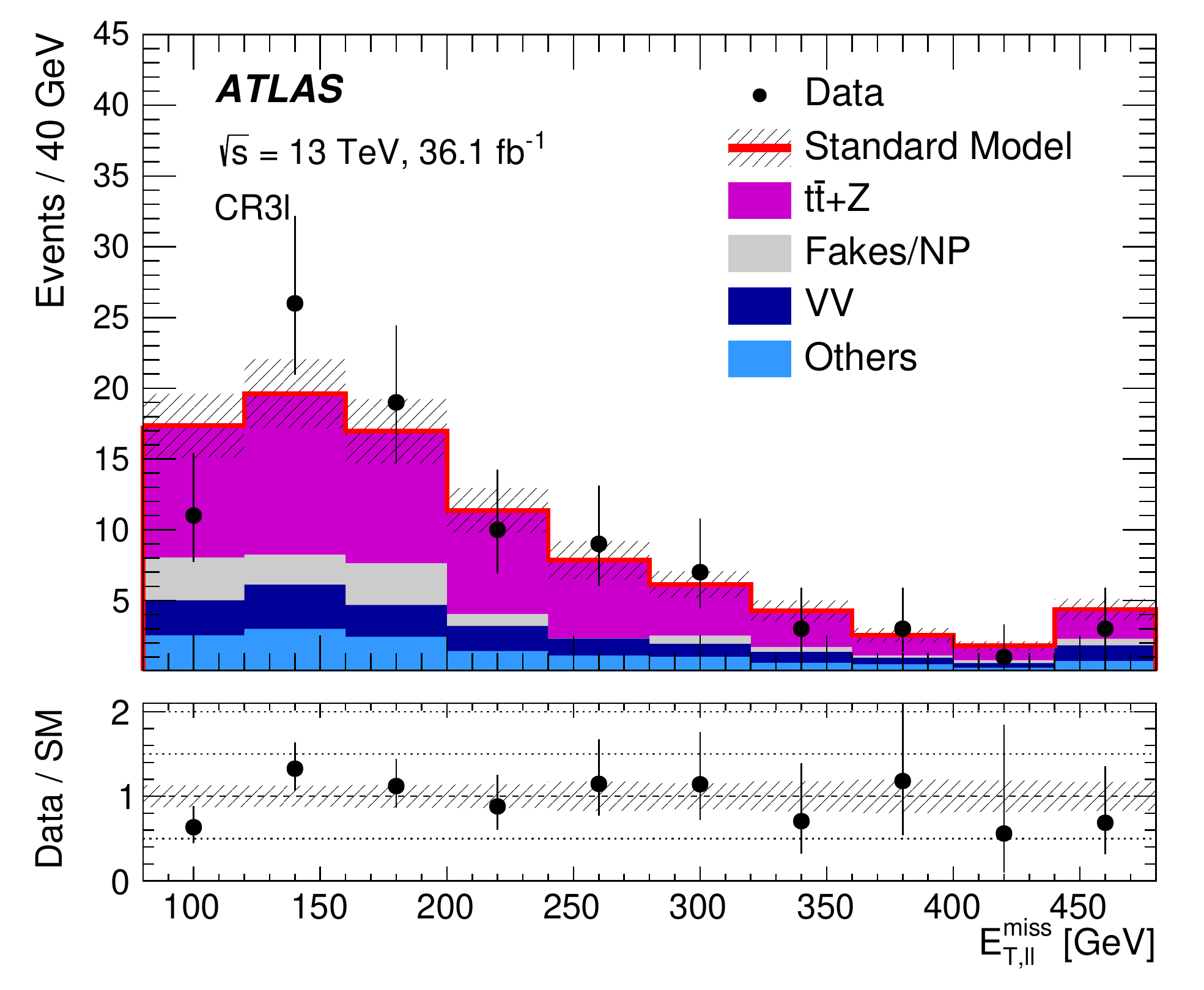}
\end{subfigure}
\caption{Comparison of the data with the post-fit Monte
Carlo prediction of some kinematic distributions in control and validation
regions. The bottom panel shows the ratio of the  data
to the Monte Carlo prediction. The band includes all systematic
uncertainties  defined in Sect.~\ref{sec:systematics}. The last bins include overflows, where applicable.
The top left
panel shows the \metprime\ distribution in CRZ\apponeb. The \metprime\
requirement is relaxed to $100\;\GeV$. The other panels show the
\cthbb\ distribution in VR\apptwob\ (top right),
the \remass{1}{1.2} distribution in VRT\appthigh\ (middle left), the \cem\
distribution the VRT\appttwol\ (middle right), the \metprime\
distribution in \crgamma\ (bottom left) and the \metprime\ distribution
in \crthreel\ (bottom right).}
\label{g:CRZs-metprime}
\end{figure}

\interfootnotelinepenalty=10000

\begin{figure}
\centering
\includegraphics[width=\textwidth]{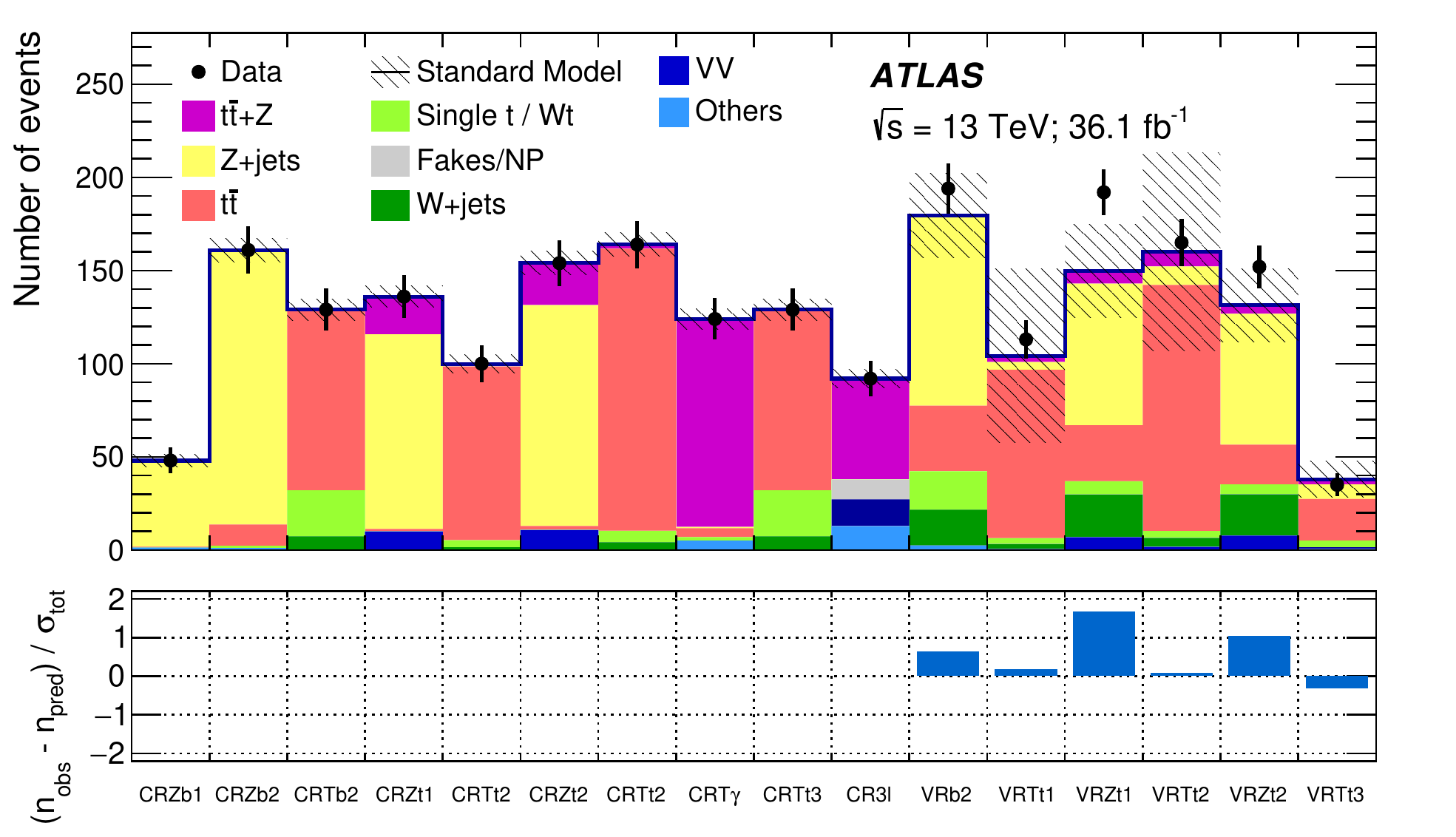}
\caption{Comparison of the data with the post-fit SM prediction of the background in each control and validation region.
The different background components are denoted by the colour specified in the legend.
All systematic uncertainties defined in Sect.~\ref{sec:systematics} and statistical uncertainties are included in the shaded band.  The lower panel shows the pulls in each VR.
The total uncertainty $\sigma_{\hbox{tot}}$ includes systematic and Poisson uncertainties for each given region.}
\label{g:VRpullplot}

\end{figure}

\noindent
A second control region (\crthreel), is used for the background normalisation
of \srttwol. It makes use of \ttZ\ events with $\Zboson\rightarrow\ell^+\ell^-$ and semileptonic
decays of the \ttbar\ system ($e$ or $\mu$). \crthreel\ is obtained by
selecting three medium leptons out of which one
SFOS pair is compatible with a $Z$-boson decay.
This strategy allows the modelling of  the lower \met\ part of the \srttwol\
signal region. Additionally, the momenta of the leptons compatible
with the \Zboson-boson decay are added vectorially to the \ptmiss\ to
define \ptmissprime\ and \metprime\ for this control region. The
transverse mass of the \ptmissprime\ and the lepton not associated with the \Zboson-boson decay, \mtprime,
is combined with the \metprime\ to define a corrected \cem: $\cemprime = \mtprime + 0.2\cdot\metprime$.
A requirement is placed on this variable in \crthreel\ in order to approximate
the kinematic properties of the signal region.
The \minbl\ variable is redefined in this region (\minblprime) as the smaller of the two transverse masses
calculated when combining the lepton not associated with the $Z$-boson decay and each of the
two \btagged\ jets in the event\footnote{When the \btagged\ jet multiplicity is different from two, the
two jets with the highest $b$-tagging probabilities are chosen, independently of whether they are
\btagged\ or not.}.
All CR selections are summarised in Table~\ref{t:CRcuts}.

The relatively small contamination of \srttwol\ and \crthreel\ from events with fake
or non-prompt (NP) leptons is estimated from data with a method similar to
that described in Refs.~\cite{TOPQ-2010-01,TOPQ-2011-01}.
Different processes contribute to this background for the two selections.
The dominant fake or non-prompt lepton contribution for \srttwol\ comes
from semileptonic \ttbar\ and \Wjets\ processes, while for \crthreel\
it comes from dileptonic \ttbar\ and $Z$+$bb$ processes.
The method makes use of the number of observed events
containing baseline--baseline, baseline--medium, medium--baseline and medium--medium
lepton pairs (see definitions in Sec.~\ref{sec:detector}) in a given
selection.
The probability for prompt leptons satisfying the baseline selection
criteria to also pass the medium selection is measured using a
$\Zboson \rightarrow \ell\ell$ sample. The equivalent
probability for fake or non-prompt leptons is measured from
multi-jet- and \ttbar-enriched control samples.
The number of events containing a contribution from one or
two fake or non-prompt leptons is calculated from these probabilities.

The background from multi-jet production for the regions with no leptons
is estimated from data using a
procedure described in detail in Ref.~\cite{SUSY-2015-01} and modified to
account for the heavy flavour of the jets. The
contribution from multi-jet production in all regions is found to
be very small.

Minor background contributions to each signal region are collectively called
"Others" in the following.
For \sroneb\ and \srtwob, this category contains the contributions from
multi-jet, single top-quark production, diboson
production, \ttV\ and \Wjets.
For \srtlow\ and \srthigh, multi-jet, $V+\gamma$, diboson, single top-quark
and \ttbar\ production in association with Higgs or \Wboson\ boson(s)
collectively define "Others".
Finally, for \srttwol\ the "Others" category contains the
contributions from \ttbar+$W/h/WW$, \ttbar\ttbar, $\ttbar t$, $Wh$, $(gg)h$ and $Zh$ production.

\begin{table}
\centering
\caption{Summary of the validation region selections. See Tables~\ref{t:srbselections} and~\ref{t:srtselections} for the detailed multi-jet rejection requirements. }
\label{t:VRcuts}
\begin{tabular}{lcccccccccc}
\toprule
Observable & VR\apptwob & VRZ\apptlow & VRZ\appthigh & VRT\apptlow & VRT\appthigh & VRT\appttwol\\
\midrule
Trigger&  \met     & \multicolumn{2}{c}{\met}&  \multicolumn{2}{c}{\met}   & $\scriptstyle 2\mu || 2e || 1e1\mu $ \\
$\begin{array}{l}\hskip-4pt\mathcal{N}_{j} \\ \hskip-4pt\mathcal{N}_{b} \end{array}$ &
$\begin{array}{c} 2\hbox{--}3 \\  \geq2 \end{array}$ &
\multicolumn{2}{c}{$\begin{array}{c} \geq4 \\ \geq2 \end{array}$} &
\multicolumn{2}{c}{$\begin{array}{c} \geq4 \\ \geq2 \end{array}$} &
$\begin{array}{c} \geq1 \\  \geq1 \end{array}$ \\
$\mathcal{N}_{\ell}$ &  \multicolumn{6}{c}{as SR} \\
$\tau$ multiplicity & - & \multicolumn{2}{c}{-}& \multicolumn{2}{c}{0} & - \\
\midrule
\met\ [\GeV]         &$>180$  & \multicolumn{2}{c}{$>250$}  & \multicolumn{2}{c}{$>300$} & -\\
$\pt(j_1,j_2)$ [\GeV] & $>150,>20$ & \multicolumn{2}{c}{$>80,>80$}& \multicolumn{2}{c}{$>80,>80$} &$>30,$- \\
$\pt(j_3,j_4)$ [\GeV] & $<60,$-    & \multicolumn{2}{c}{$>40,>40$}& \multicolumn{2}{c}{$>40,>40$} & - \\
$\pt(bj_1)$ [\GeV] & $>150$ & \multicolumn{2}{c}{$>20$} & \multicolumn{2}{c}{$>20$}& $>30$ \\
$\pt(\ell_1,\ell_2)$ [\GeV] & - & \multicolumn{2}{c}{-}& \multicolumn{2}{c}{-} & $>25,20$ \\
Multi-jet rejection &  \multicolumn{6}{c}{as SR} \\
$\vert\mll^{\mathrm{SF}} \hskip-3pt - m_\Zboson\vert$\ [\GeV] & - & \multicolumn{2}{c}{-}& \multicolumn{2}{c}{-} & $>20$\\
\midrule
\xone, \yone\ [rad] & $<0, >0.5$ &   \multicolumn{2}{c}{-}&   \multicolumn{2}{c}{-}& - \\[0.13ex]
\remass{0}{SR} [\GeV] & - & $<80$& $<140$& $>80$& $>140$ & - \\[0.13ex]
\remass{1}{SR} [\GeV] & - &  \multicolumn{2}{c}{-} & $>40$& $>50$ & - \\[0.13ex]
\mtbmin\ [\GeV] &  - & \multicolumn{2}{c}{$>150$} & $(80,150)$ & $(100,200)$ & - \\[0.13ex]
\mtbmax\ [\GeV] &  -  & $>250$ & - & $>200$    & - & - \\[0.13ex]
\drbb\  &  - &  \multicolumn{2}{c}{$<1.5$}&  $>0.8$ & $>1.0$ & - \\[0.13ex]
\metsig [$\sqrt{\GeV}$] &  - & $>12$ & - & - & $>10$ & - \\[0.13ex]
\cem, \minbl, \mttwoll\ [\GeV] &  - &  \multicolumn{2}{c}{-} &  \multicolumn{2}{c}{-} & as SR \\[0.13ex]
\dphib\ [rad] &  - &  \multicolumn{2}{c}{-} &  \multicolumn{2}{c}{-} & $>1.5$ \\[0.13ex]
\bottomrule
\end{tabular}
\end{table}

In summary, one scaling factor is used to normalise the \Zjets\ background in \sroneb,
while two scaling factors are used to normalise the \Zjets\ and \ttbar\ backgrounds in \srtwob.
For \srtlow\ and \srthigh, three scaling factors for each region are used to independently normalise the
\Zjets, \ttbar\ and \ttZ\ backgrounds. Finally, in \srttwol\ the \ttbar\ and \ttZ\ predictions are adjusted
by a floating normalisation for each of the two backgrounds.
The background scaling factors are treated as fully uncorrelated between the different SRs.
In all selections, it is found that the normalisation of the \Zjets\ background is larger than unity.
This may be related to the fact that in the default \SHERPA v2.2.1 generator the heavy-flavour production fractions
are not consistent with the measured values~\cite{ATL-PHYS-PUB-2017-007}.
The normalisation factors for \ttbar\ processes in the SRtX regions are found to be compatible with unity, while
they are found to be considerably smaller than unity for \srtwob. This is due to the angular separation requirements in
this region, which select \ttbar\ events in a specific corner of the phase space. Finally, the different normalisations of the \ttZ\
background processes found in the \crgamma\ and \crthreel\
regions (larger and smaller than unity, respectively) are due to the different kinematic requirements on the jet momenta
and the corrected \met\ in the two regions, which are designed to mimic the topology of the respective signal regions.

Dedicated validation regions are used to validate the background prediction for each of the SRs and evaluate the reliability of the MC extrapolation of the SM background estimates from CRs to SRs.
The background estimates in \srtwob\ are validated in a single VR (VR\apptwob) which has a
background composition similar to that of the SR.
Selected key distributions in the control and validation regions are shown in Fig.~\ref{g:CRZs-metprime}.
The prediction of the \Zjets\ background in \sroneb\ relies on an extrapolation
over a large interval of missing transverse momentum.
As CRZ\apponeb\ is designed to be kinematically as close as possible to \sroneb\ and
given the low yield in this region, it was not possible to construct a selection
to validate this extrapolation.
Nevertheless, the use of the same kinematic selection in control and signal region, together with the good
agreement between the data and the post-fit SM prediction in CRZ\apponeb\  in the
whole \metprime\ spectrum  (Fig.~\ref{g:CRZs-metprime}) gives confidence in the accuracy of the estimate.
Two validation regions, VRZ\apptlow\ and VRZ\appthigh, are designed to validate the \Zjets\ estimate in
\srtlow\ and \srthigh. Furthermore, the top background estimate in these SRs is validated in two additional VRs: VRT\apptlow\ and VRT\appthigh.
Finally, VRT\appttwol\ is designed to validate the top background prediction in \srttwol.
All requirements for each validation region are summarised in Table~\ref{t:VRcuts}.
The  data and the post-fit Monte Carlo background prediction yields in each CR and VR are compared in
Fig.~\ref{g:VRpullplot}. The background yields in the control regions match the observed data by construction.
In the validation regions, the
background prediction is compatible with the observed data within two standard
deviations of the total systematic uncertainty.

\FloatBarrier
\section{Systematic uncertainties}
\label{sec:systematics}
Experimental and theoretical sources of systematic uncertainty in the signal and background estimates are considered in this analysis.
Their impact is constrained overall through the normalisation of the dominant backgrounds in the control regions defined with kinematic selections resembling those of the corresponding signal region.

The dominant sources of detector-related systematic uncertainty are the jet energy scale, the jet energy resolution, the $b$-tagging efficiency and mis-tagging rates,
and the scale and resolution of the \MET\ soft term.
The jet energy scale and resolution uncertainties are derived as a
function of the \pT and $\eta$ of the jet, as well as of the pile-up
conditions
and the jet flavour composition of the selected jet sample~\cite{ATL-PHYS-PUB-2015-015}.
Uncertainties associated with the modelling of the $b$-tagging
efficiencies for $b$-jets, $c$-jets and light-flavour
jets~\cite{ATLAS-CONF-2014-004,ATLAS-CONF-2014-046} are derived as a
function of
$\eta$, \pt and flavour of each jet.
The systematic uncertainties related to the modelling of \MET\ in the simulation are estimated by propagating the uncertainties in the energy and momentum scale of all identified electrons, photons, muons and jets, as well as the uncertainties in the soft-term scale and resolution~\cite{ATL-PHYS-PUB-2015-023}.
Other detector-related systematic uncertainties, such as those in the lepton and photon reconstruction efficiency, energy scale and energy resolution, and in
the modelling of the trigger~\cite{PERF-2015-10}, are found to have a small impact on the results.

\begin{table}
\caption{
Summary of the main systematic uncertainties and their impact  on the
total SM background prediction in each of the signal regions studied. A range is shown for the four bins composing \srtwob . The
total systematic uncertainty can be different from the sum in quadrature of
individual uncertainties due to the correlations between them resulting from the fit
to the data.
\label{tab:syst}
}
\begin{center}
\setlength{\tabcolsep}{0.0pc}
{\small
\begin{tabular*}{0.9\textwidth}{@{\extracolsep{\fill}}lrrrrr}
\noalign{\smallskip}\hline\noalign{\smallskip}
&  \sroneb\ [\%]&  \srtwob\  [\%]& \srtlow\ [\%]& \srthigh\ [\%]&\srttwol\  [\%]\\

\noalign{\smallskip}\hline\noalign{\smallskip}
Total systematic uncertainty   & 18 & 15--18 & 29 & 14 & 28  \\
\noalign{\smallskip}\hline\noalign{\smallskip}
\Zboson theoretical uncertainties  & 5.7 & 7.9--12  & 5.0 &  2.1 & <1 \\
\ttZ\ theoretical uncertainties    &  <1 & <1    & 3.3 & 5.3 & 8.4 \\
\ttbar theoretical uncertainties   & <1    & 2.7--9.8 & 17 & 5.7 & 11 \\
\noalign{\smallskip}\hline\noalign{\smallskip}
MC statistical uncertainties       & 6.4 & 4.8--6.4 &15 & 5.9 & 18\\
\noalign{\smallskip}\hline\noalign{\smallskip}
\Zboson\ fitted normalisation         &  13 & 12--19   &  2.3   & 3.4 & - \\
\ttZ\ fitted normalisation        &  - & -    & 2.2  & 3.5 & 7.1 \\
\ttbar fitted normalisation       & -    & 1.9--4.2      & 3.9 & 1.4  & 2.0 \\
\noalign{\smallskip}\hline\noalign{\smallskip}
Fake or non-prompt leptons        & -   & - & -    & -   & 7.9 \\
Pile-up                    & 3.8  & <1--1.4   & 6.8 & 5.5 & <1\\
Jet energy resolution       & 1.5 & 1.3--6.9  & 7.0 & <1& <1 \\
Jet energy scale        & 7.7 &5.0--10  &5.0 & 2.8 &  8.2 \\
\MET soft term            & <1 & 4.3--6.3  & 2.0 & <1 & 12\\
$b$-tagging            & <1  & 2.4--6.9 & 8.6 & 3.1 & <1 \\

\noalign{\smallskip}\hline\noalign{\smallskip}
\end{tabular*}
}
\end{center}

\end{table}

Uncertainties in the theoretical modelling of the SM background processes from MC simulation are also taken into account.
The uncertainties in the modelling of the $\ttbar$ process are estimated by varying
the renormalisation and factorisation scales, as well as the amount of initial- and final-state radiation used to generate the samples ~\cite{ATL-PHYS-PUB-2016-004}.
The uncertainty connected with the parton-shower modelling is estimated as the difference between the predictions from \POWHEG showered with \PYTHIA or \HERWIG.
Additionally, the  uncertainty related to the choice of event generator is evaluated by comparing
the \POWHEG and \AMCATNLO predictions~\cite{ATL-PHYS-PUB-2016-004} for \sroneb, \srtwob\ and \srttwol.
Due to the higher jet multiplicity required in \srtlow\ and \srthigh\  the generator uncertainty is evaluated instead by comparing the \POWHEG and \SHERPA predictions.
The uncertainties in the modelling of the \Zboson background are accounted for by varying the default renormalisation, factorisation, resummation and matching scales of the \SHERPA samples.
For \srtlow\ and \srthigh\ an additional uncertainty is included
to account for effects on the \drbb\ modelling not captured by the scale variations.
This is estimated as the difference between the observed yield in data and the post-fit background prediction plus one times its uncertainty in each of the VRZs.
The theoretical uncertainty connected with the $\ttbar Z$ background in \srtlow\ and \srthigh\ is estimated by varying independently the renormalisation, factorisation, resummation and matching scales
in the $\ttbar Z$ and $\ttbar \gamma$ samples in signal and control regions, respectively. PDF uncertainties (estimated by varying the parametrisation of the PDF set used to generate the simulated background samples) are found to have a non-negligible impact for this
background component and are treated as correlated between signal and control regions.
An additional uncertainty in the extrapolation between control and signal region is derived as the difference between the ratio of the $\ttbar \gamma$ and $\ttbar Z$ cross-section
predictions obtained with the nominal MC generator and with the alternative MC generator \SHERPA interfaced to \OPENLOOPS.
For \srttwol , \sroneb\ and \srtwob\ the uncertainty connected with the $\ttbar Z$ background estimation is assessed by varying the
renormalisation, factorisation, resummation and matching scales.

Systematic uncertainties are assigned to the estimated background from fake or non-prompt leptons in \srttwol\  to account for potentially different compositions
(heavy flavour, light flavour or conversions) between the signal
regions and the control regions used for the fake-rate extraction, as well as the contamination
from prompt leptons in the regions used to measure the probabilities for loose fake or non-prompt leptons to satisfy the tight signal criteria.
Table~\ref{tab:syst} summarises the contributions from the different sources of systematic uncertainty
in the total SM background predictions for the different signal regions after the fit to the control regions described in Sect.~\ref{sec:bkg}.
As can be seen, the contribution from the theoretical uncertainty in the $\ttbar$ background
and the contribution from the statistical uncertainty connected with the use of Monte Carlo simulations are higher in \srtlow\ than in \srthigh .
The reason for the higher contribution from the theoretical uncertainty in the $\ttbar$ background is primarily due to the larger relative importance of this source of background in \srtlow .
The reason for the higher contribution from the statistical uncertainty is connected with the $W$-boson background, which is predicted with low statistical precision in \srtlow .

The impact of  theoretical and detector-related uncertainties on
the dark-matter signal acceptance is considered.
The same procedure used to evaluate background uncertainties is applied for the detector-related uncertainties.
The theoretical uncertainties in the acceptance are
assessed by varying the factorisation, renormalisation, matching scales and parton shower parameters.
For \sroneb\ the total theoretical uncertainty in the acceptance is $6$\%, for \srtwob\  it is below $8$\%, and for  \srtlow, \srthigh\
and \srttwol\ it ranges from $10$\% to $12$\%.
The theoretical uncertainties in the production cross-section of the
signal are evaluated only for the colour-neutral mediator models, for
which an NLO computation of the cross-section is available. It is estimated by considering
the same scale variations used to assess the uncertainties in the
acceptance,
and by varying the parametrisation of the PDF set used to generate the simulated signal samples.
An additional uncertainty due to the different scale adopted to evaluate the NLO cross-section
and to generate the signal samples is also considered.
The total theoretical uncertainty in the cross-section amounts to $9$\%
for the on-shell regime in the mass range of $\ttbar+\phi/a$ signals to which
the analysis is sensitive, and ranges from $9$\% to $30$\% for the off-shell
regime. For the $\bbbar+\phi/a$ signals this uncertainty varies
between $5$\% and $13$\%.

\FloatBarrier

\section{Results}
\label{sec:result}
\begin{table}[t]
\caption{Fit results in \sroneb\ and \srtwob\ for an integrated luminosity of \intlumi. The background normalisation
parameters are obtained from the background-only fit in the CRs and are applied to the SRs. Pre-fit values are also shown.
Small backgrounds are indicated as Others (see text for details). The dominant component of these smaller background sources in \sroneb\ is diboson processes.
Benchmark signal models yields are given for each SR. The uncertainties in the yields include statistical uncertainties and all systematic uncertainties  defined in Sect.~\ref{sec:systematics}.}
\label{t:SRyields1}
\begin{center}
{\small
\begin{tabular*}{\textwidth}{@{\extracolsep{\fill}}lrrrrr}
\toprule
&       \sroneb    & \srtwob-bin1            & \srtwob-bin2            & \srtwob-bin3            & \srtwob-bin4              \\[-0.05cm]
\midrule
Observed      &  $19$    & $88$              & $88$              & $90$              & $82$                    \\[0.5ex]
\midrule
Total background (fit)    & $16.9 \pm 3.3$ & $77 \pm 13$          & $72 \pm 11$          & $76 \pm 13$          & $66.4 \pm 9.1$              \\[0.5ex]
\noalign{\smallskip}\hline\noalign{\smallskip}
$Z/\gamma^* \text{+ jets}$& $14.2 \pm 3.1$         & $39.7 \pm 6.3$       & $44.4 \pm 6.6$      & $53.3 \pm 9.9$ & $55.6 \pm 8.6$              \\[0.5ex]
\ttbar         & $0.58_{-0.58}^{+0.60}$    & $17.8 \pm 6.5$       & $13.8 \pm 5.5$      & $14.0 \pm 4.7$ & $7.0 \pm 2.9$              \\[0.5ex]
Single top quark     & $0.25_{-0.25}^{+0.42}$    & $14.7 \pm 5.8$       & $10.2 \pm 3.7$      & $5.5 \pm 3.1$  & $2.6 \pm 1.7$              \\[0.5ex]
Others        & $2.0 \pm 1.1$                  & $5.2 \pm 3.4$                  & $3.4_{-1.6}^{+1.7}$   & $2.7 \pm 1.1$    & $1.3 \pm 1.0$              \\[0.5ex]
\midrule
$Z/\gamma^* \text{+ jets}$ (pre-fit) & $12.1$         & $30.6$          & $34.2$          & $41.1$    & $42.8$              \\[0.5ex]
\ttbar    (pre-fit)     & -   & $27.1$          & $21.1$          & $21.4$    & $10.6$              \\[0.5ex]
\midrule
Signal benchmarks\\
\midrule
$m(\phi,\chi)=(20,1)\;\GeV$,  $g=1$ & & $0.238\pm 0.085$&   $0.262\pm 0.079$&   $0.320\pm 0.082$&   $0.277\pm 0.080$\\[0.5ex]
$m(a,\chi)=(20,1)\;\GeV$,  $g=1$    & & $0.256\pm 0.065$&    $0.199\pm 0.060$&   $0.308\pm 0.085$&   $0.267\pm 0.067$\\[0.5ex]
$m(\phi_b,\chi)=(1000,35)\;\GeV$  & $18.6\pm3.8$& &  & &\\[0.5ex]
\bottomrule
\end{tabular*}
}
\end{center}
\end{table}

\begin{table}[p]
\caption{Fit results in \srtlow, \srthigh\ and \srttwol\ for an integrated luminosity of \intlumi. The background normalisation
parameters are obtained from the background-only fit in the CRs and are applied to the SRs. Pre-fit values are also shown.
Small backgrounds are indicated as Others (see text for details). Benchmark signal models yields are given for each SR. The uncertainties in the yields include statistical uncertainties and all systematic uncertainties defined in Sect.~\ref{sec:systematics}.}
\label{t:SRyields2}
\begin{center}
\setlength{\tabcolsep}{0.1pc}
{\small
\begin{tabular*}{\textwidth}{@{\extracolsep{\fill}}lrrr}
\toprule
& \srtlow & \srthigh & \srttwol              \\[-0.05cm]
\midrule
Observed         & $23$&$24$   & $18$                    \\[0.5ex]
\noalign{\smallskip}\hline\noalign{\smallskip}
Total background (fit)      & $20.5\pm5.8$  & $20.4\pm2.9$    & $15.2 \pm 4.3$              \\[0.5ex]
\midrule
$t\bar{t}$                     & $7.0\pm3.9$  & $3.1\pm1.3$   & $4.5 \pm 2.5$              \\[0.5ex]
$t\bar{t}$+$Z$                 & $4.3\pm1.1$  & $6.9\pm1.4$   & $4.4 \pm 1.9$              \\[0.5ex]
\Wjets                         & $3.3\pm2.6$  & $1.28\pm0.50$    & incl. in Fakes/NP\\[0.5ex]
$Wt$                           &   incl. in Others & incl. in Others & $0.33_{-0.33}^{+0.53}$              \\[0.5ex]
$Z/\gamma^* \text{+ jets}$    & $3.7\pm1.4$  & $6.2\pm1.1$   &      incl. in Others       \\[0.5ex]
$VV$                             & incl. in Others & incl. in Others  & $0.61 \pm 0.25$              \\[0.5ex]
Fakes/NP       & - & -  & $2.7 \pm 1.3$              \\[0.5ex]
Others                         & $2.2\pm1.2$          & $3.00\pm1.6$            & $2.69 \pm 0.93$              \\[0.5ex]
\midrule
$t\bar{t}$ (pre-fit)                     & $6.1 $  & $2.8$   & $4.0$    \\[0.5ex]
$t\bar{t}$+$Z$  (pre-fit)               & $3.53$  & $5.6$   & $5.6$    \\[0.5ex]
$Z/\gamma^* \text{+ jets}$  (pre-fit)   & $3.2$    & $5.72$   & -    \\[0.5ex]
\midrule
Signal benchmarks\\
\midrule
$m(\phi,\chi)=(20,1)\;\GeV$,  $g=1$   &  $9.3 \pm 1.6$  &       $12.8 \pm 1.9$   & $21.0 \pm 2.3$  \\[0.5ex]
$m(a,\chi)=(20,1)\;\GeV$,  $g=1$       &  $7.6 \pm 1.5$ &        $12.1 \pm 1.8$  &  $14.1 \pm 1.6$ \\[0.5ex]
$m(\phi,\chi)=(100,1)\;\GeV$,  $g=1$  & $6.5 \pm 1.3$ &        $10.1 \pm 1.5$  & $11.5 \pm 1.5$\\[0.5ex]
$m(a,\chi)=(100,1)\;\GeV$,  $g=1$      &  $6.2 \pm 1.2$ &   $11.5 \pm 2.0$  & $11.9 \pm 1.5$  \\[0.5ex]
\bottomrule
\end{tabular*}
}
\end{center}
\end{table}

\begin{table}[p]
\caption{Left to right: 95\% CL upper limits on the visible cross-section
($\langle\epsilon\mathcal{A}\sigma\rangle^{\rm obs}_{95}$) and on the number of
BSM events ($S^{\rm obs}_{95}$ ).  The third column
($S^{\rm exp}_{95}$) shows the 95\% CL upper limit on the number of
signal events, given the expected number (and $\pm 1\sigma$
excursions of the expected number) of background events.
The last column
indicates
the discovery $p$-value ($p(s = 0)$) and $Z$ (the number of equivalent Gaussian standard deviations).}
\label{t:BSMres}
\centering
\setlength{\tabcolsep}{0.0pc}
\begin{tabular*}{\textwidth}{@{\extracolsep{\fill}}lcccc}
\noalign{\smallskip}\hline\noalign{\smallskip}
\toprule
{\bf Signal channel}                        &
$\langle\epsilon\mathcal{A}{\rm \sigma}\rangle^{\rm obs}_{95}$[fb]  &  $S^{\rm obs}_{95}$  & $S^{\rm exp}_{95}$ & $p(s=0)$ ($Z$) \\
\midrule
\sroneb            & $0.37$ &  $13.4$ & $ { 12 }^{ +5 }_{ -1 }$   & $ 0.33$~$(0.43)$ \\
\srtwob\  bin-1    & $1.10$ &  $39.6$ & $ { 33 }^{ +12 }_{ -8 }$  & $ 0.22$~$(0.76)$ \\
\srtwob\  bin-2   & $1.17$ &  $42.1$ & $ { 31 }^{ +10 }_{ -8 }$   & $ 0.11$~$(1.21)$ \\
\srtwob\  bin-3   & $1.21$ &  $43.7$ & $ { 33 }^{ +11 }_{ -8 }$  & $ 0.16$~$(1.00)$ \\
\srtwob\  bin-4    & $1.10$ &  $39.8$ & $ { 26 }^{ +11 }_{ -7 }$  & $ 0.10$~$(1.26)$ \\
\srtlow           & $0.51$ &  $18.4$ & $ { 16 }^{ +5 }_{ -4 }$   & $ 0.33$~$(0.44)$ \\
\srthigh           & $0.44$ &  $15.7$ & $ { 12 }^{ +5 }_{ -3 }$   & $ 0.24$~$(0.70)$ \\
\srttwol          & $0.44$ &  $15.9$ & $ { 13 }^{ +5 }_{ -2 }$   & $ 0.33$~$(0.45)$ \\
\bottomrule
\end{tabular*}
\end{table}

The expected and observed yields in each of the five signal regions of this analysis are
reported in Tables~\ref{t:SRyields1}~and~\ref{t:SRyields2}.  The background-only
fit to the control regions described in Sect.~\ref{sec:bkg} is
compared to the predictions based on the MC normalisation. The observed data is found to be compatible with
the background prediction in each one of the SRs.  The
expected signal yields for selected benchmark models
for colour-neutral and colour-charged mediators are also shown.
In each SR the observed yield in data is above the expected background
but within 1.3 standard deviations of its uncertainty.

\begin{figure}
\centering
\centering
\begin{subfigure}{.48\textwidth}\centering
\includegraphics[width=.98\textwidth]{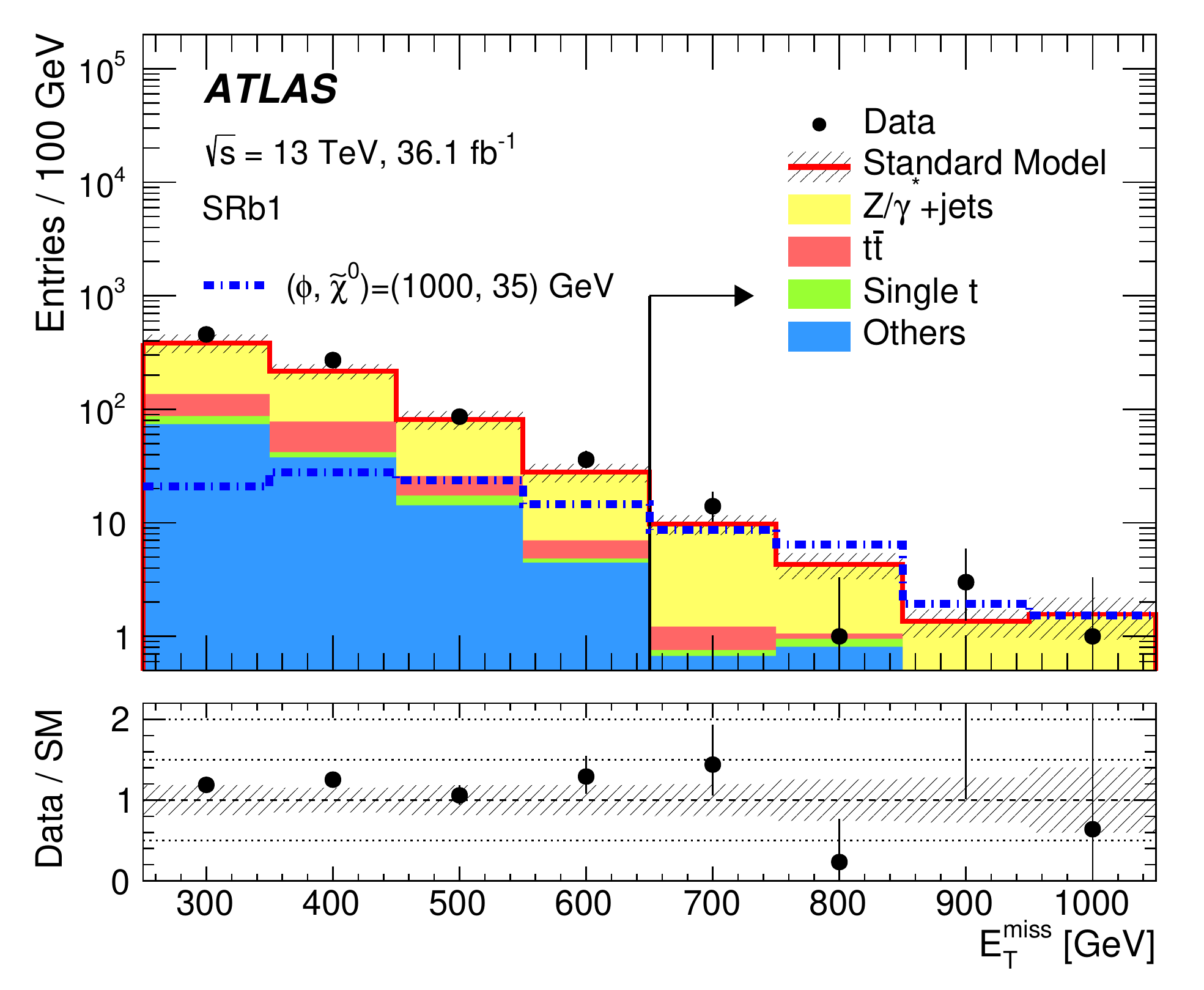}
\end{subfigure}
\begin{subfigure}{.48\textwidth}\centering
\includegraphics[width=.98\textwidth]{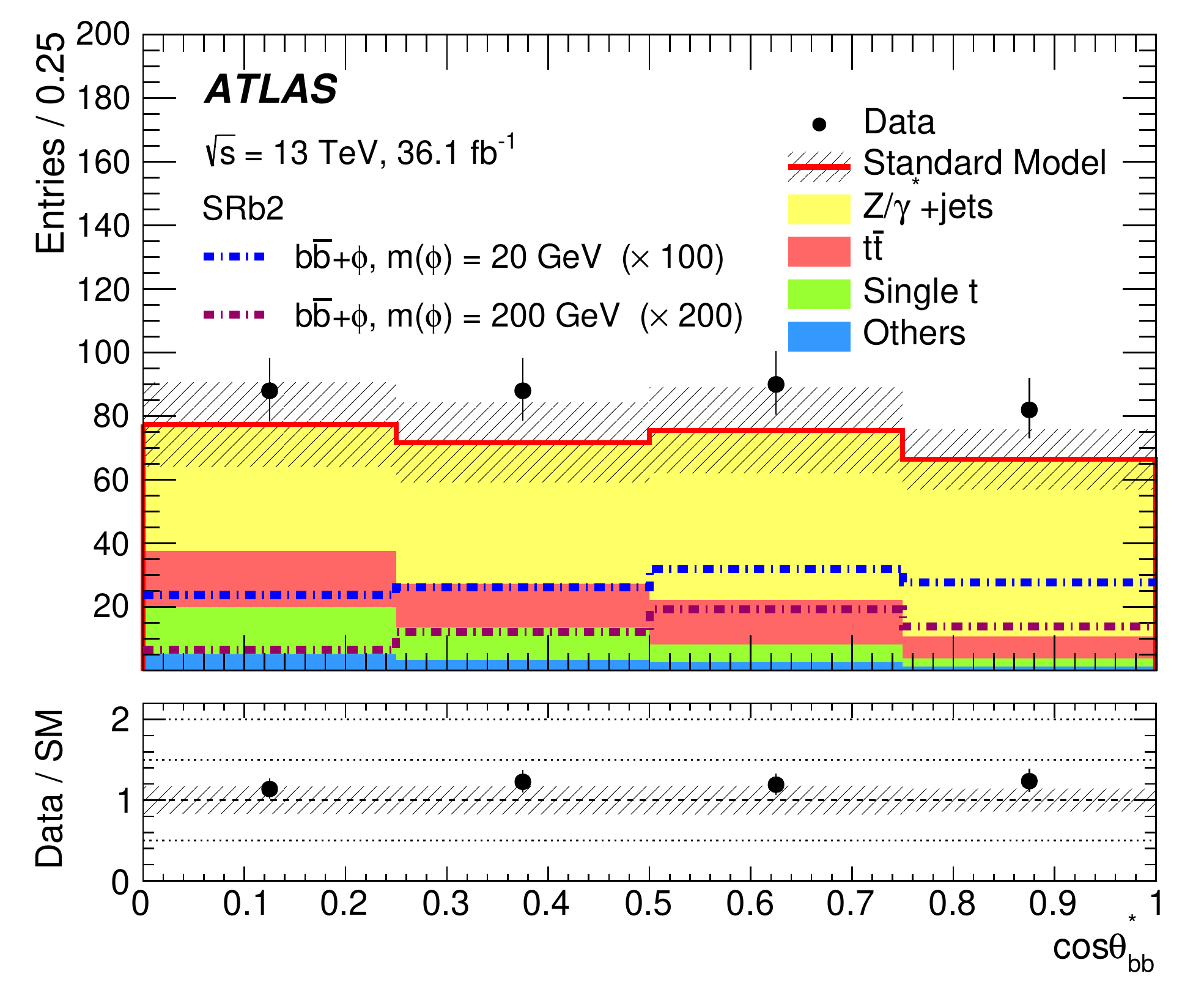}
\end{subfigure}

\begin{subfigure}{.48\textwidth}\centering
\includegraphics[width=.98\textwidth]{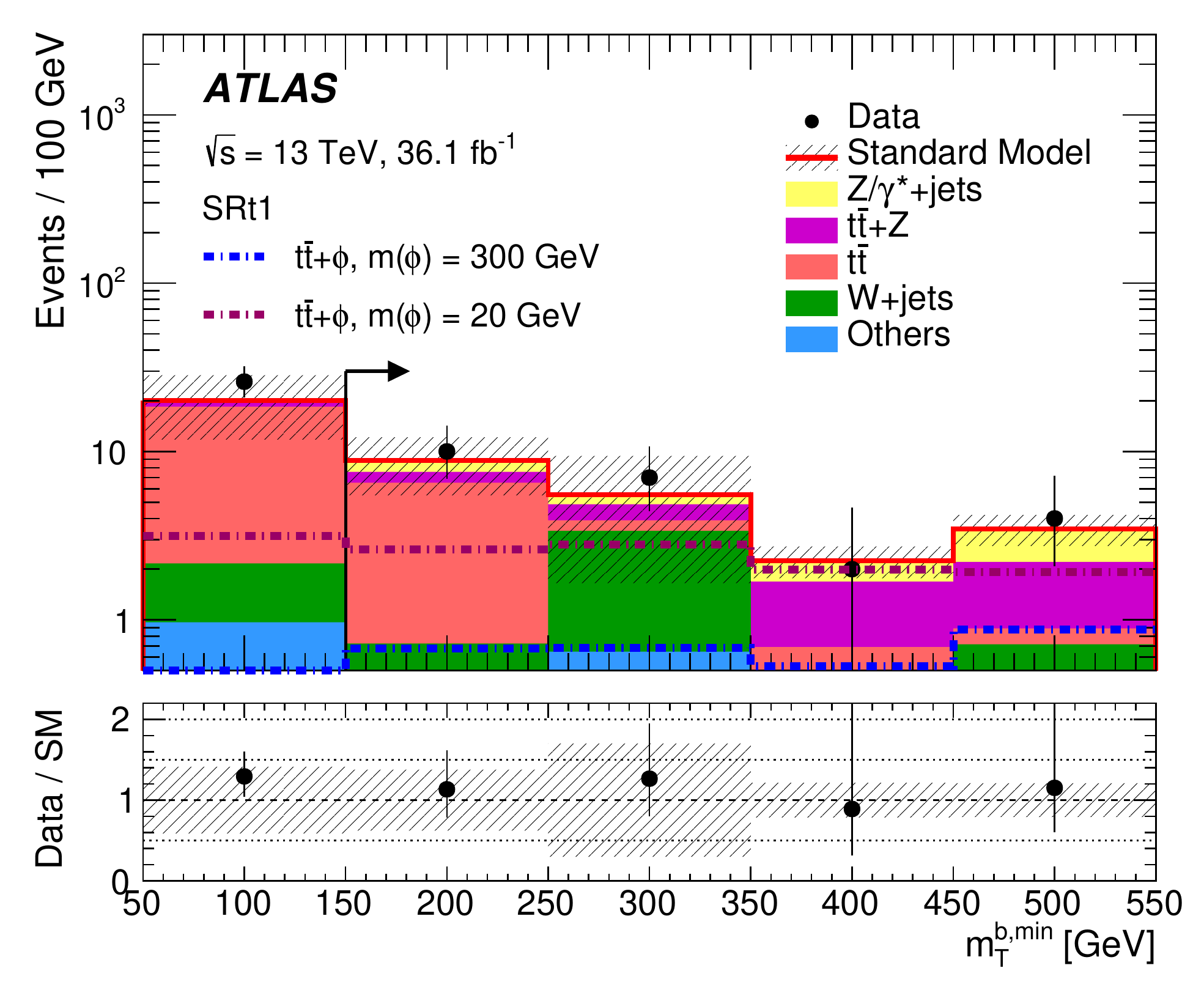}
\end{subfigure}
\begin{subfigure}{.48\textwidth}\centering
\includegraphics[width=.98\textwidth]{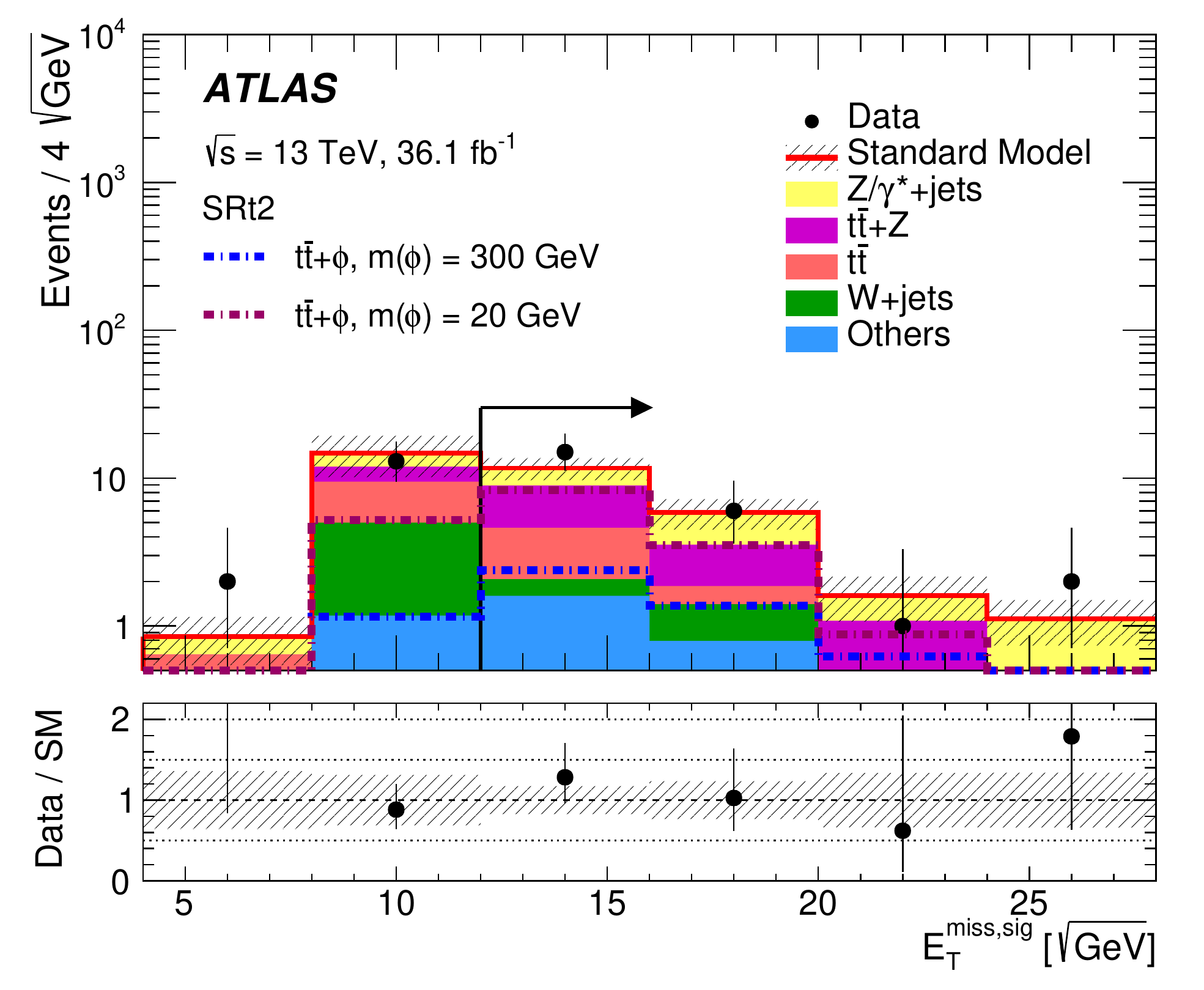}
\end{subfigure}

\begin{subfigure}{.48\textwidth}\centering
\includegraphics[width=.98\textwidth]{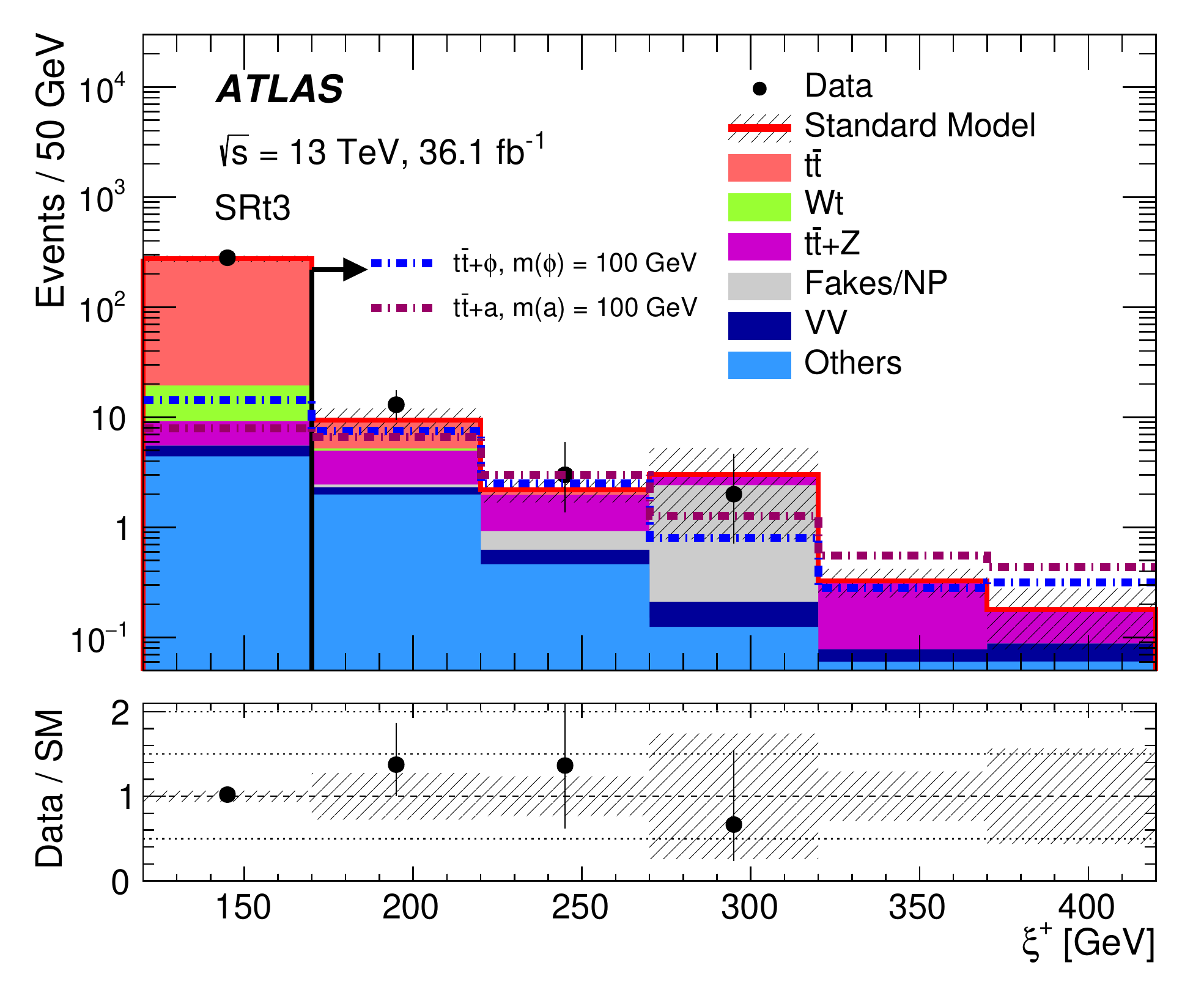}
\end{subfigure}
\caption{Comparison of the  data with the post-fit SM prediction of the \met\ distribution in  \sroneb\ (top left), \cthbb\ distribution in \srtwob\ (top right), \mtbmin\ distribution in \srtlow\ (middle left), \metsig\ distribution in \srthigh\ (middle right) and \cem\ distribution in \srttwol\ (bottom). The last bins include overflows, where applicable. All signal region requirements except the one on the distribution shown are applied.
The signal region requirement on the distribution shown is indicated by an arrow. The bottom panel shows the ratio of the  data to the prediction. The band includes all systematic uncertainties  defined in Sect.~\ref{sec:systematics}.}
\label{fig:SRNm1}
\end{figure}

Figure~\ref{fig:SRNm1}  shows a comparison between the SM
predictions and the observed data for some relevant kinematic
distributions in each signal region prior to the selection on the variable.
The four bins of \srtwob\ are
statistically combined in the final result.
A model-independent fit set-up~\cite{Baak:2014wma} where both
the control and signal regions are included in the fit is used to derive
95\% confidence level (CL) upper limits on the visible cross-section $\langle\epsilon\mathcal{A}{\rm \sigma}\rangle_{95}$ of new physics beyond-the-SM (BSM) processes, defined as cross-section times
acceptance times efficiency and obtained as the upper limit on the number of
BSM events divided by the total integrated luminosity. The 95\% CL exclusion limits are derived with the
CL$_\mathrm{s}$~method~\cite{ReadCLs} and summarised in
Table~\ref{t:BSMres} for each SR.
These limits are calculated assuming no systematic uncertainties for
the signal and neglecting any possible
signal contamination in the control regions.

The results are also used to set limits on the production cross-section of
colour-neutral and colour-charged mediator models decaying into dark-matter
particles.
An independent fit is used for each of the five signal regions.
When deriving model-dependent limits, the expected signal yield in each
fit region is considered.

For the signal, the experimental
systematic uncertainties and theoretical systematic uncertainties in the acceptance are taken into account for this calculation.
The experimental uncertainties are assumed to be
fully correlated with those in the SM background.
The theoretical systematic uncertainties in the signal cross-section are instead shown separately in the final exclusion result
for the colour-neutral mediator models.

\begin{figure}
\centering
\begin{subfigure}{.9\textwidth}\centering
\includegraphics[width=\textwidth]{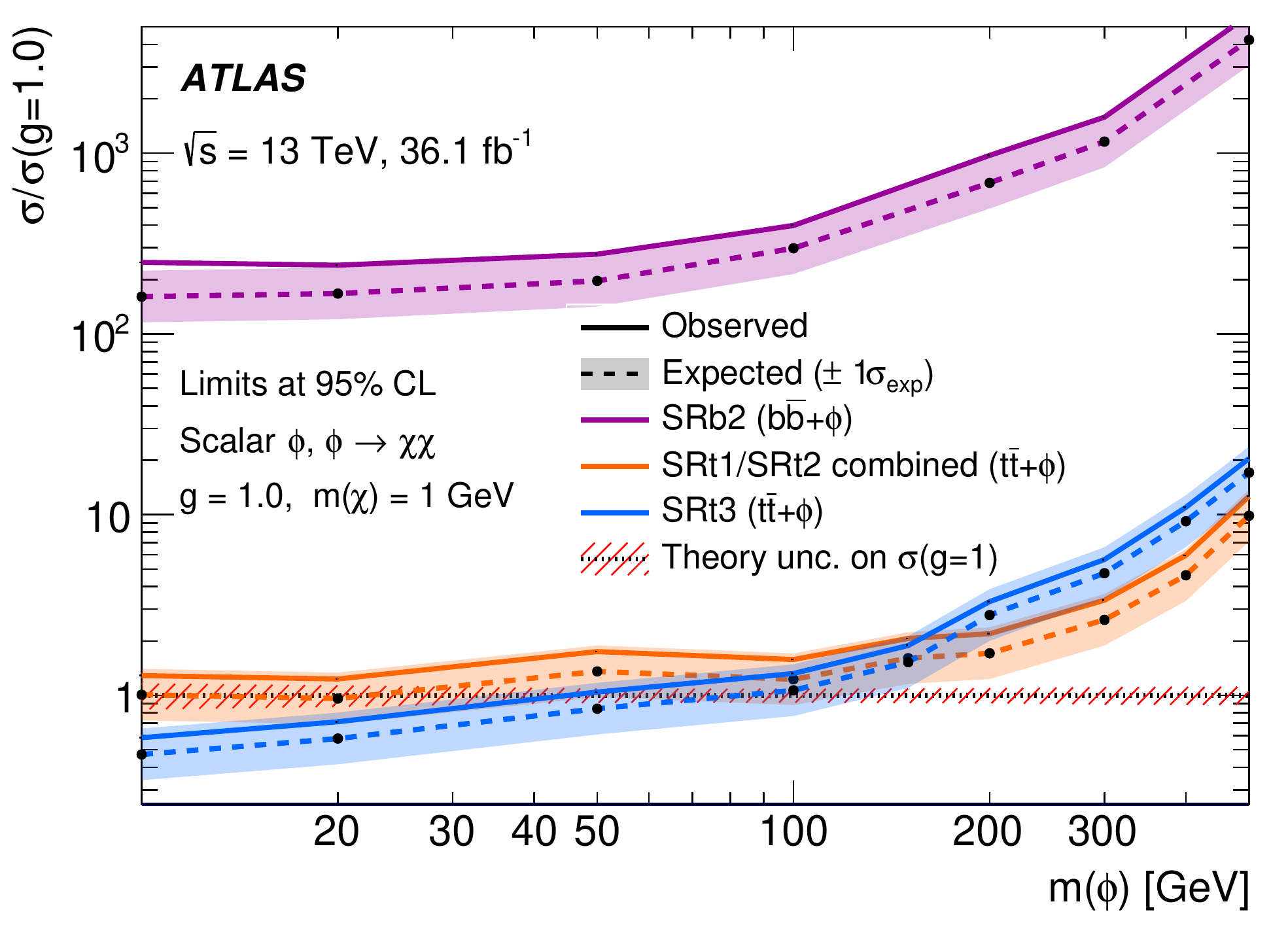}
\end{subfigure}

\begin{subfigure}{.9\textwidth}\centering
\includegraphics[width=\textwidth]{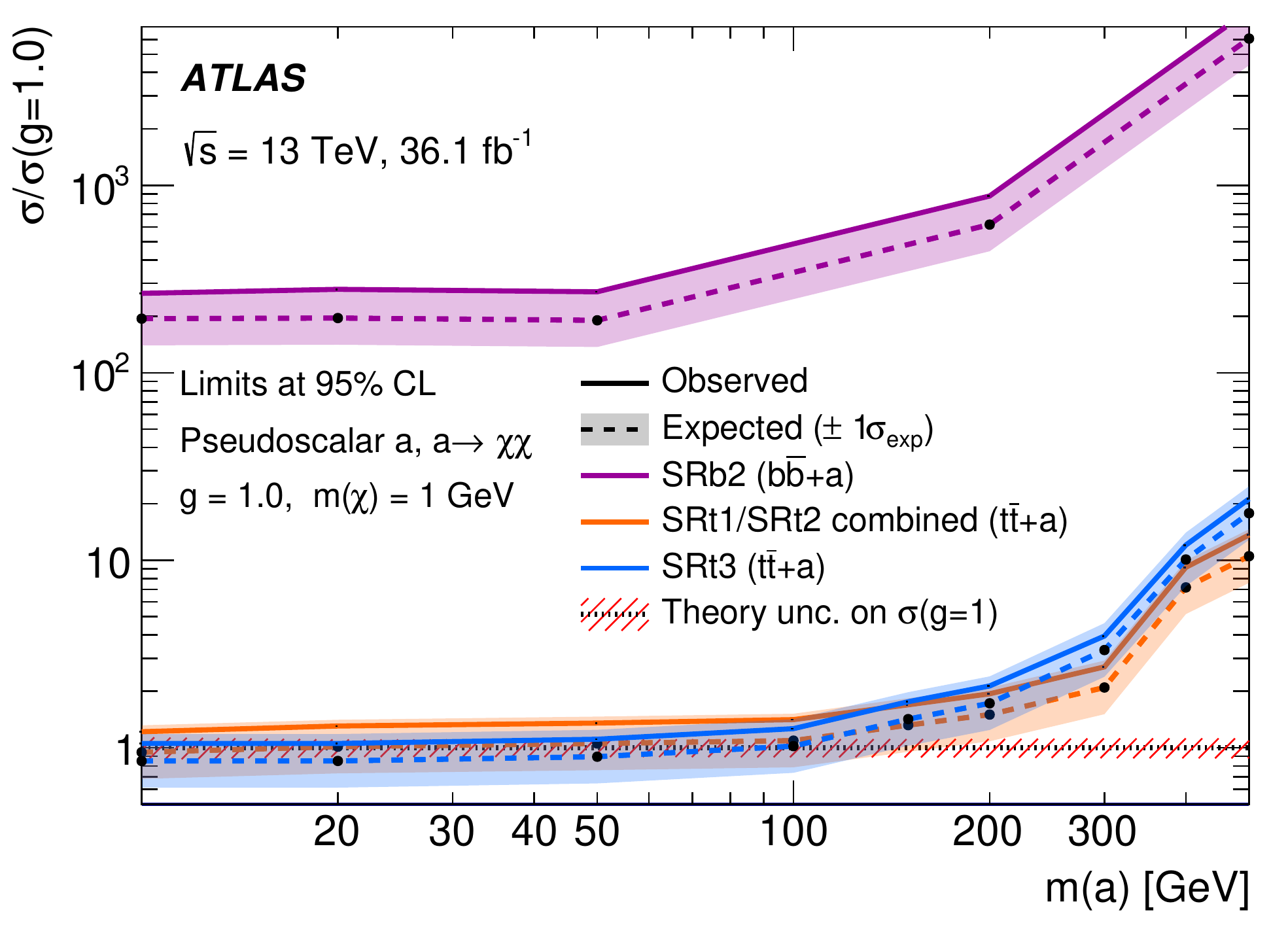}
\end{subfigure}
\caption{Exclusion limits for colour-neutral $\ttbar/\bbbar+\phi$  scalar (top) and $\ttbar/\bbbar+a$ pseudoscalar (bottom)  models as a function of
the mediator mass for a DM mass of $1\;\GeV$.
The limits are calculated at 95\% CL and are expressed in terms of
the ratio of the excluded cross-section to the nominal cross-section
for a coupling assumption of $g = g_q = g_\chi = 1$. The solid (dashed) lines shows the observed (expected) exclusion limits for the different signal regions, according to the colour code specified in the legend.  To derive the results for the fully hadronic \ttbar\ final state the region \srtlow\ or \srthigh\ providing the better expected sensitivity is used.
}
\label{fig:Results}\centering
\end{figure}

\begin{figure}
\centering
\begin{subfigure}{.9\textwidth}\centering
\includegraphics[width=\textwidth]{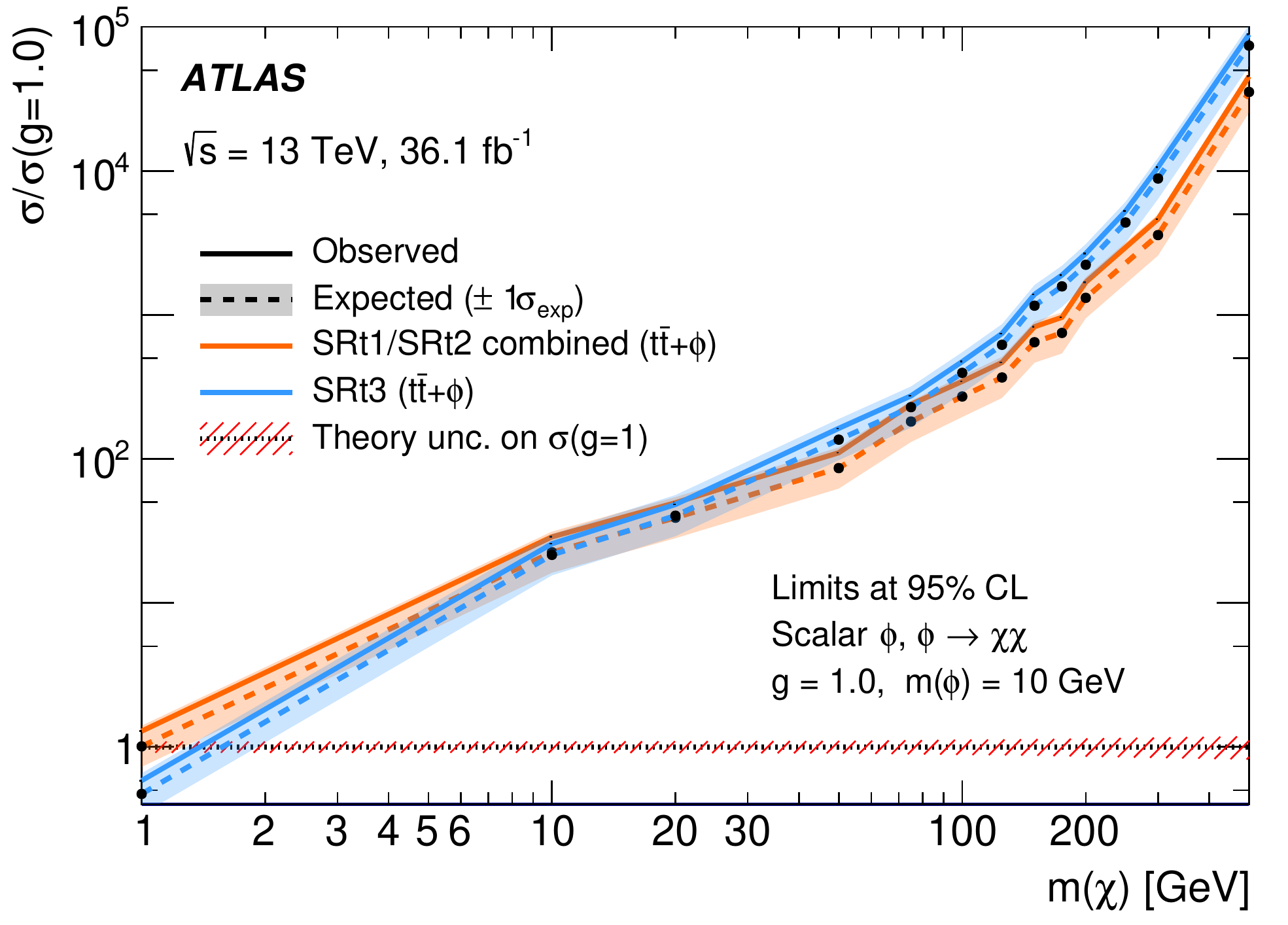}
\end{subfigure}

\begin{subfigure}{.9\textwidth}\centering
\includegraphics[width=\textwidth]{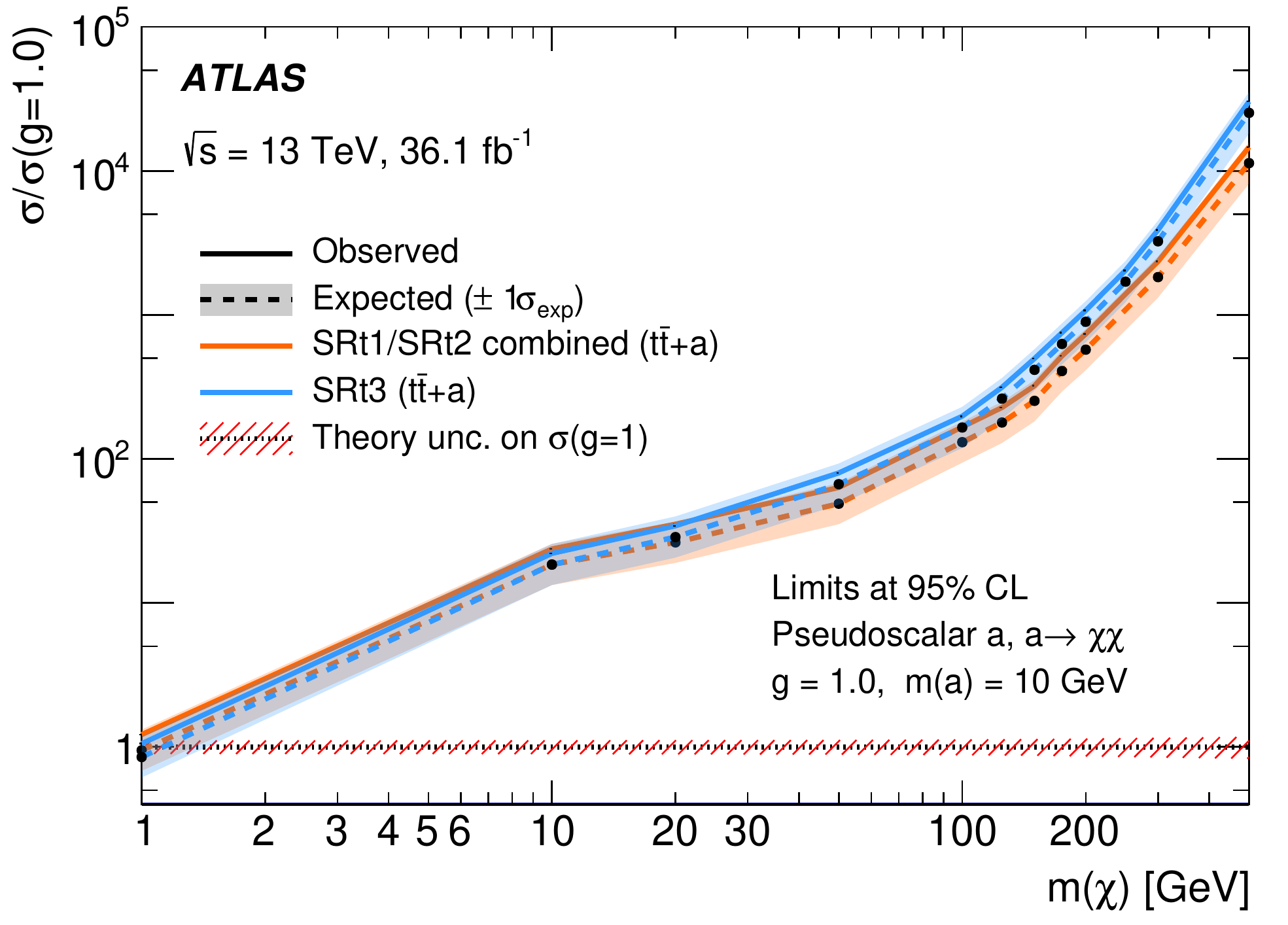}
\end{subfigure}
\caption{Exclusion limits for colour-neutral $\ttbar+\phi$ scalar (top) and $\ttbar+a$ pseudoscalar (bottom)  models
as a function of the
DM mass for a mediator mass of $10\;\GeV$.
The limits are calculated at 95\% CL and are expressed in terms of
the ratio of the excluded cross-section to the nominal cross-section
for a coupling assumption of $g = g_q = g_\chi = 1$. The solid (dashed) lines shows the observed (expected) exclusion limits for the different signal regions, according to the colour code specified in the legend. To derive the results for the fully hadronic \ttbar\ final state the region \srtlow\ or \srthigh\ providing the better expected sensitivity is used.
}
\label{fig:Resultschi}\centering
\end{figure}

Figures~\ref{fig:Results}~and~\ref{fig:Resultschi}
show upper limits at 95\% CL
on the signal cross-section scaled to the signal cross-section for
coupling
$g=1$, denoted by $\sigma/\sigma(g=1.0)$.
These are the most stringent limits to date on $\ttbar+\phi/a$ models
and the first ATLAS results for the $\bbbar+\phi/a$ models.
To derive the results for the fully hadronic \ttbar\ final state the
region \srtlow\ or \srthigh\ providing the better expected
sensitivity is used. The \srtlow\ was originally optimised for
low-mass scalar mediators, while \srthigh\ was optimised for high-mass
scalar mediators and pseudoscalar mediators. However, \srtlow\ is
strongly affected by systematic uncertainties in the
\ttbar\ modelling and therefore \srthigh\ sets more stringent limits
for the whole parameter space.
These limits are obtained both as a function of the mediator mass,
assuming a specific DM mass of $1\; \GeV$ (Fig.~\ref{fig:Results}), and as a function of the
DM mass, assuming a specific mediator mass of $10\;\GeV$ (Fig.~\ref{fig:Resultschi}).
Both the scalar and pseudoscalar mediator cases are considered.
The sensitivity for $\ttbar+\phi/a$ on-shell decays is approximately constant for masses
below $100\;\GeV$, with \srttwol\ excluding the $g=1$ assumption for
scalar mediator masses up to  $50\;\GeV$.
For a given mediator mass the acceptance of the analysis is independent
of the value of the DM mass as long as $m(\phi /a) > 2\cdot m(\chi)$ is fulfilled and width effects can be neglected.
Under these conditions, exclusion limits for DM masses differing from the one presented can be inferred from the result shown in Fig.~\ref{fig:Results}.
Due to the smaller Yukawa enhancement
of $\bbbar+\phi/a$ final states, it is possible to exclude cross-sections $300$ times the nominal
values for $g=1$.

\begin{figure}[p]
\centering
\includegraphics[width=.56\textwidth]{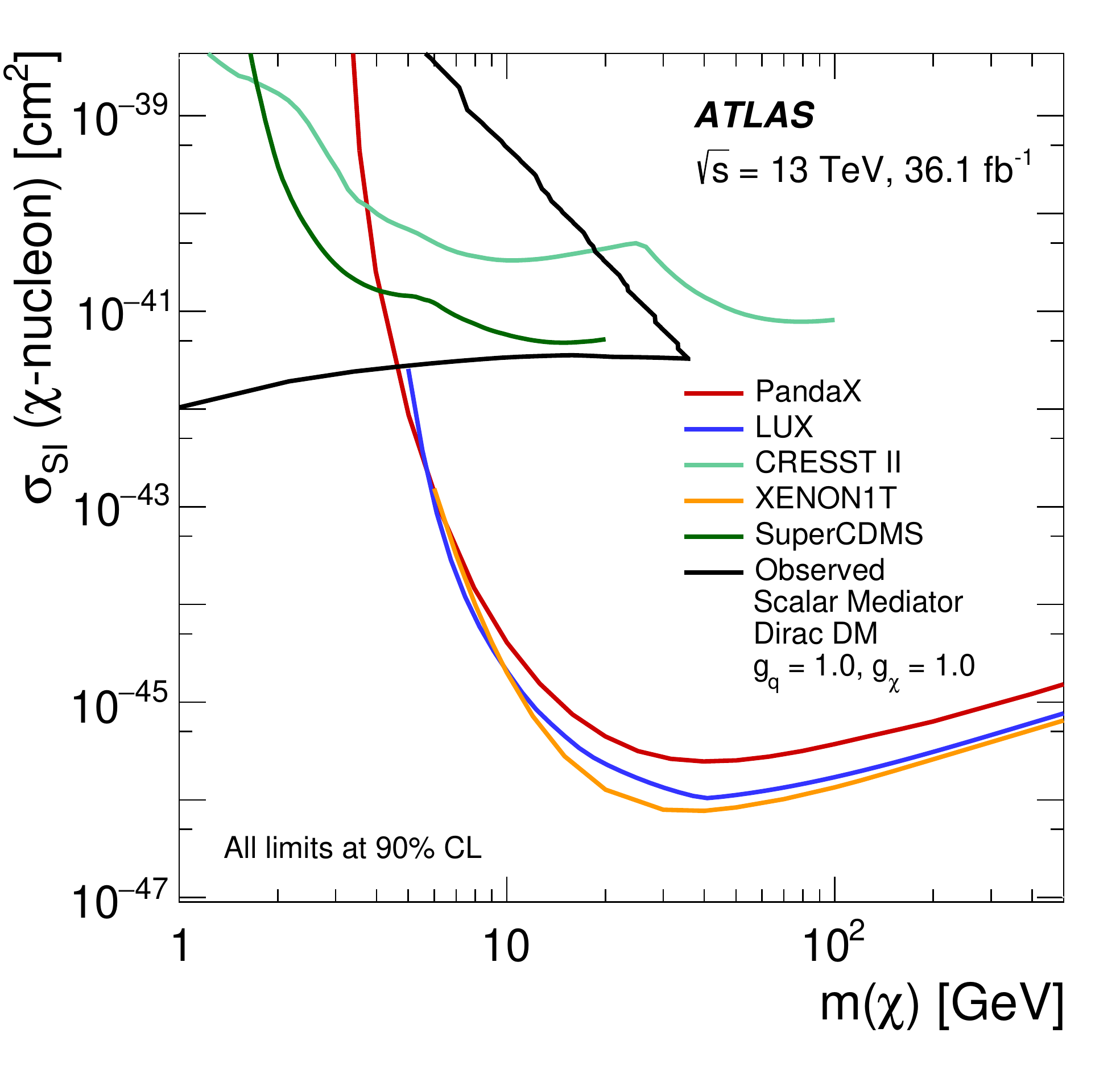}
\caption{
Comparison of the 90\% CL limits on the spin-independent DM--nucleon cross-section as a function of DM mass between these results and the direct-detection experiments,
in the context of the colour-neutral simplified model with scalar mediator.
The black line indicates the exclusion contour
derived from the observed limits of \srttwol. Values inside the contour are excluded.
The exclusion limit is compared with limits from the
LUX~\cite{Lux}, PandaX-II~\cite{Panda},
XENON~\cite{Xenon}, SuperCDMS~\cite{SuperCDMS} and CRESST-II~\cite{CRESST}
experiments.
}
\label{fig:DDlimit}
\end{figure}

\begin{figure}[p]
\centering
\includegraphics[width=.72\textwidth]{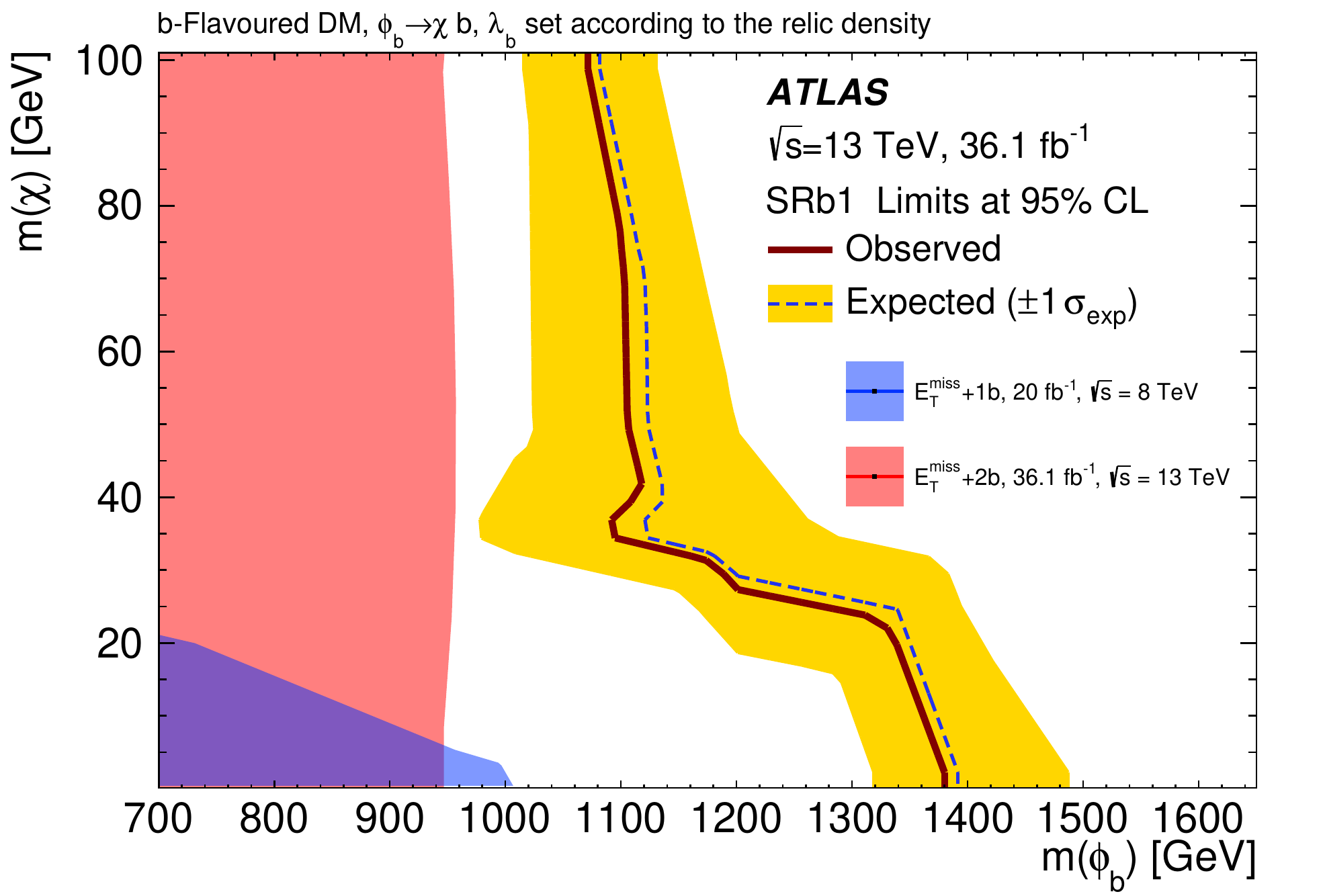}
\caption{Exclusion limits for colour-charged scalar mediators (\bFDM) as a
function of the mediator and DM masses for \intlumi\ of data.  The
limits are calculated at 95\% CL. The solid (dashed) line show the observed (expected) exclusion contour
for a coupling assumption $\lambda_b$ yielding the measured relic density. No uncertainties on the LO cross-sections are considered for this model.
The results are compared with the ATLAS search for \bFDM\ models \cite{Aad:2014vea}, represented by the blue contour, and the ATLAS search for direct sbottom pair production \cite{Sbottom}, represented by the red contour.}
\label{fig:Results2}
\end{figure}

For each dark-matter and mediator mass pair, the exclusion limit on the production cross-section of
colour-neutral scalar mediator particles can be converted into a limit on the spin-independent DM--nucleon
scattering cross-section using the procedure described in Ref.~\cite{CERN-LPCC-2016-001}.
The results can thus be compared with the results from direct-detection experiments.
The most stringent limits, provided by \srttwol , are used for this purpose.
Figure~\ref{fig:DDlimit} shows the constraints from this analysis expressed as exclusion limits at
90\% CL  in the plane defined by the dark-matter mass and the scattering cross-section.
The black line indicates the exclusion contour derived from the observed limits in the top part of
Fig.~\ref{fig:Results}, where mediator masses between 10 GeV and 500 GeV are considered.
The maximum value of the DM--nucleon scattering cross-section displayed
corresponds to the result obtained for a mediator mass of 10 GeV.
The results of this analysis are compared with the results from the LUX~\cite{Lux}, PandaX-II~\cite{Panda},
XENON~\cite{Xenon}, SuperCDMS~\cite{SuperCDMS} and CRESST-II~\cite{CRESST}
experiments. The comparison is model-dependent, and therefore valid only for the specific models considered in this paper.
For pseudoscalar mediator models, the predicted dark-matter cross-sections in these direct-detection experiments is suppressed by velocity-dependent terms.
As a result, direct-detection limits on spin-independent DM--nucleon
scattering cross-section are several orders of magnitude worse than the ones obtained in this analysis, and therefore not presented.

Finally, Fig.~\ref{fig:Results2} shows the exclusion contour
for the \bFDM\ model as a function of the mediator and DM masses.
In this model, the cross-section and therefore also the final sensitivity
strongly depends on the coupling choice, $\lambda_b$, which is set to fulfil
the relic density constraints, and determines the decrease of the sensitivity
for higher DM masses.
For a DM particle of approximately $35\; \GeV$, as suggested by the
interpretation of data recorded by the Fermi-LAT Collaboration,
mediator masses below $1.1\; \TeV$ are
excluded at 95\% CL.

\FloatBarrier

\section{Conclusion}
\label{sec:conclusion}
This article reports  a search for dark-matter
pair production in association with bottom or top quarks.
The analysis is performed using  \intlumi\ of $pp$ collisions collected
at a centre-of-mass energy of $\sqrt{s} = 13~\TeV$ by the ATLAS detector at the LHC.
The results are interpreted
in the framework of simplified models of spin-0 mediators
to the dark sector decaying into pairs of DM particles. The data are found to
be consistent with the Standard Model expectations, and limits are set on the
signal strength for a coupling assumption of $g=1.0$ or on the DM and mediator masses.
The results represent the most stringent limits to date for colour-neutral spin-0 mediator models for a
DM mass assumption of $1\;\GeV$ in top-quark final states.
It excludes at 95\% CL mediator masses between
$10$ and $50\; \GeV$\ for scalar mediators assuming couplings equal to unity and a dark-matter mass of $1\;\GeV$.
Although the analysis is expected to be sensitive to models with pseudoscalar mediators with masses between 10 and 100 GeV, no observed exclusion limit can be set
for this model for the coupling assumption of $g=1.0$ because of a small excess in the observed data.
Limits of $300$ times the nominal cross section for
couplings equal to unity are placed for scalar and pseudoscalar mediator masses between $10$ and $50\; \GeV$
for a dark-matter mass of $1\;\GeV$ in bottom-quark final states.
Constraints on \bFDM\ models are also presented.
The excluded region depends on $m(\phi_b)$ and $m(\chi)$;
for $m(\chi) = 35\; \GeV$, mediator particles with
$m(\phi) < 1.1\; \TeV$ are excluded.

\section*{Acknowledgements}
%

We thank CERN for the very successful operation of the LHC, as well as the
support staff from our institutions without whom ATLAS could not be
operated efficiently.

We acknowledge the support of ANPCyT, Argentina; YerPhI, Armenia; ARC, Australia; BMWFW and FWF, Austria; ANAS, Azerbaijan; SSTC, Belarus; CNPq and FAPESP, Brazil; NSERC, NRC and CFI, Canada; CERN; CONICYT, Chile; CAS, MOST and NSFC, China; COLCIENCIAS, Colombia; MSMT CR, MPO CR and VSC CR, Czech Republic; DNRF and DNSRC, Denmark; IN2P3-CNRS, CEA-DSM/IRFU, France; SRNSF, Georgia; BMBF, HGF, and MPG, Germany; GSRT, Greece; RGC, Hong Kong SAR, China; ISF, I-CORE and Benoziyo Center, Israel; INFN, Italy; MEXT and JSPS, Japan; CNRST, Morocco; NWO, Netherlands; RCN, Norway; MNiSW and NCN, Poland; FCT, Portugal; MNE/IFA, Romania; MES of Russia and NRC KI, Russian Federation; JINR; MESTD, Serbia; MSSR, Slovakia; ARRS and MIZ\v{S}, Slovenia; DST/NRF, South Africa; MINECO, Spain; SRC and Wallenberg Foundation, Sweden; SERI, SNSF and Cantons of Bern and Geneva, Switzerland; MOST, Taiwan; TAEK, Turkey; STFC, United Kingdom; DOE and NSF, United States of America. In addition, individual groups and members have received support from BCKDF, the Canada Council, CANARIE, CRC, Compute Canada, FQRNT, and the Ontario Innovation Trust, Canada; EPLANET, ERC, ERDF, FP7, Horizon 2020 and Marie Sk{\l}odowska-Curie Actions, European Union; Investissements d'Avenir Labex and Idex, ANR, R{\'e}gion Auvergne and Fondation Partager le Savoir, France; DFG and AvH Foundation, Germany; Herakleitos, Thales and Aristeia programmes co-financed by EU-ESF and the Greek NSRF; BSF, GIF and Minerva, Israel; BRF, Norway; CERCA Programme Generalitat de Catalunya, Generalitat Valenciana, Spain; the Royal Society and Leverhulme Trust, United Kingdom.

The crucial computing support from all WLCG partners is acknowledged gratefully, in particular from CERN, the ATLAS Tier-1 facilities at TRIUMF (Canada), NDGF (Denmark, Norway, Sweden), CC-IN2P3 (France), KIT/GridKA (Germany), INFN-CNAF (Italy), NL-T1 (Netherlands), PIC (Spain), ASGC (Taiwan), RAL (UK) and BNL (USA), the Tier-2 facilities worldwide and large non-WLCG resource providers. Major contributors of computing resources are listed in Ref.~\cite{ATL-GEN-PUB-2016-002}.

%

\printbibliography
\newpage

\begin{flushleft}
{\Large The ATLAS Collaboration}

\bigskip

M.~Aaboud$^\textrm{\scriptsize 137d}$,
G.~Aad$^\textrm{\scriptsize 88}$,
B.~Abbott$^\textrm{\scriptsize 115}$,
O.~Abdinov$^\textrm{\scriptsize 12}$$^{,*}$,
B.~Abeloos$^\textrm{\scriptsize 119}$,
S.H.~Abidi$^\textrm{\scriptsize 161}$,
O.S.~AbouZeid$^\textrm{\scriptsize 139}$,
N.L.~Abraham$^\textrm{\scriptsize 151}$,
H.~Abramowicz$^\textrm{\scriptsize 155}$,
H.~Abreu$^\textrm{\scriptsize 154}$
% ...
\bigskip
\\
% ...
$^{12}$ Institute of Physics, Azerbaijan Academy of Sciences, Baku, Azerbaijan\\
% ...
$^{88}$ CPPM, Aix-Marseille Universit{\'e} and CNRS/IN2P3, Marseille, France\\
% ...
$^{115}$ Homer L. Dodge Department of Physics and Astronomy, University of Oklahoma, Norman OK, United States of America\\
% ...
$^{119}$ LAL, Univ. Paris-Sud, CNRS/IN2P3, Universit{\'e} Paris-Saclay, Orsay, France\\
% ...
$^{137}$ $^{(a)}$ Facult{\'e} des Sciences Ain Chock, R{\'e}seau Universitaire de Physique des Hautes Energies - Universit{\'e} Hassan II, Casablanca; $^{(b)}$ Centre National de l'Energie des Sciences Techniques Nucleaires, Rabat; $^{(c)}$ Facult{\'e} des Sciences Semlalia, Universit{\'e} Cadi Ayyad, LPHEA-Marrakech; $^{(d)}$ Facult{\'e} des Sciences, Universit{\'e} Mohamed Premier and LPTPM, Oujda; $^{(e)}$ Facult{\'e} des sciences, Universit{\'e} Mohammed V, Rabat, Morocco\\
% ...
$^{139}$ Santa Cruz Institute for Particle Physics, University of California Santa Cruz, Santa Cruz CA, United States of America\\
% ...
$^{151}$ Department of Physics and Astronomy, University of Sussex, Brighton, United Kingdom\\
% ...
$^{154}$ Department of Physics, Technion: Israel Institute of Technology, Haifa, Israel\\
$^{155}$ Raymond and Beverly Sackler School of Physics and Astronomy, Tel Aviv University, Tel Aviv, Israel\\
% ...
$^{161}$ Department of Physics, University of Toronto, Toronto ON, Canada\\
% ...
$^{a}$ Also at Department of Physics, King's College London, London, United Kingdom\\
$^{b}$ Also at Institute of Physics, Azerbaijan Academy of Sciences, Baku, Azerbaijan\\
$^{c}$ Also at Novosibirsk State University, Novosibirsk, Russia\\
$^{d}$ Also at TRIUMF, Vancouver BC, Canada\\
$^{e}$ Also at Department of Physics {\&} Astronomy, University of Louisville, Louisville, KY, United States of America\\
% ...
$^{*}$ Deceased
\end{flushleft}


\clearpage


\appendix


\end{document}